\documentclass[twocolumn]{svjour3}          % twocolumn
\smartqed  % flush right qed marks, e.g. at end of proof
\usepackage{graphicx}
\graphicspath{{./}{./figures/}}
\usepackage{natbib}
\usepackage{url}
\usepackage{eurosym}

\begin{document}

\title{Faint objects in motion: the new frontier of high precision
  astrometry}
% \thanks{Grants or other notes about the article that should go on
% the front page should be placed here. General acknowledgments should
% be placed at the end of the article.}


\author{%
  Fabien Malbet \and
  C\'eline Boehm \and
  Alberto Krone-Martins \and
  Antonio Amorim \and
  Guillem Anglada-Escud\'e \and
  Alexis Brandeker \and
  Fr\'ed\'eric Courbin \and
  Torsten En{\ss}lin \and
  Antonio Falc\~{a}o \and
  Katherine Freese \and
  Berry Holl \and
  Lucas Labadie \and
  Alain L\'eger \and
  Gary A.\ Mamon \and                % modif name
  Barbara McArthur \and
  Alcione Mora \and
  Mike Shao \and
  Alessandro Sozzetti \and
  Douglas Spolyar \and
  Eva Villaver \and
  Ummi Abbas \and
  Conrado Albertus \and
  Jo\~{a}o Alves \and
  Rory Barnes \and
  Aldo Stefano Bonomo \and
  Herv\'e Bouy \and
  Warren R.\ Brown \and                  %% modif name
  Vitor Cardoso \and
  Marco Castellani \and
  Laurent Chemin \and
  Hamish Clark \and
  Alexandre C.\ M.\ Correia \and       %% modif name
  Mariateresa Crosta \and
  Antoine Crouzier \and
  Mario Damasso \and
  Jeremy Darling \and
  Melvyn B.\ Davies \and
  Antonaldo Diaferio \and
  Morgane Fortin \and
  Malcolm Fridlund \and
  Mario Gai \and
  % Arianna Gallo \and                                                                 % added
  Paulo Garcia \and
  Oleg Gnedin \and
  Ariel Goobar \and
  Paulo Gordo \and
  Renaud Goullioud \and
  David Hall \and
  Nigel Hambly \and
  Diana Harrison \and
  David Hobbs \and
  Andrew Holland \and
  Erik H\o{}g \and
  Carme Jordi \and
  Sergei Klioner \and
  Ariane  Lan\c{c}on \and
  Jacques Laskar \and
  Mario Lattanzi \and
  Christophe Le Poncin-Lafitte \and
  Xavier Luri \and
  Daniel Michalik \and
  Andr\'e Moitinho de Almeida \and
  Ana Mour\~{a}o \and
  Leonidas Moustakas \and
  Neil J.\ Murray \and                                                                            % modif name (?)
  Matthew Muterspaugh \and
  Micaela Oertel \and
  Luisa Ostorero \and
  Jordi Portell \and
  Jean-Pierre Prost \and
  Andreas Quirrenbach \and                                                                   % name (s)
  % Krzysztof A.\ Rybicki \and                                                                  % added
  Jean Schneider \and
  Pat Scott \and
  Arnaud Siebert \and
  Antonio da Silva \and
  Manuel Silva \and
  Philippe Th\'ebault \and
  John Tomsick \and
  Wesley Traub \and
  Miguel de Val-Borro \and 
  Monica Valluri \and
  Nicholas A.\ Walton \and
  Laura L.\ Watkins \and
  Glenn White \and
  Lukasz Wyrzykowski \and
  Rosemary Wyse \and
  Yoshiyuki Yamada
}%


\institute{%
  F.~Malbet \at Univ.\ Grenoble Alpes, CNRS, IPAG, Grenoble, France\\
  \email{fabien.malbet@univ-grenoble-alpes.fr}
  \and C.~Boehm \at The University of Sydney, School of Physics, Camperdown, Australia                  % modif affil
  \and A.~Krone-Martins \at CENTRA/University of  Lisbon, Lisbon, Portugal
  \and A.~Amorim \at FCUL, CENTRA/SIM, Lisbon, Portugal 
  \and G.~Anglada-Escud\'e \at Queen Mary University of London, UK
  \and A.~Brandeker \at Stockholm University, Sweden
  \and F.~Courbin \at EPFL, Lausanne, Switzerland
  \and T.~En{\ss}lin \at Max-Planck Institute for Astrophysics, Garching, Germany
  \and A.~Falc\~{a}o \at Uninova, Caparica, Portugal
  \and K.~Freese \at University Texas, Austin, TX, USA
                           \at Stockholm University, Sweden
  \and B.~Holl \at Dep. of Astronomy, Univ. Geneva, Switzerland                                                         %modif affil
  \and L.~Labadie \at University of  Cologne, Cologne, Germany
  \and A.~L\'eger \at Univ.\ Paris-Saclay, CNRS, Institut d'astrophysique spatiale, Orsay, France
  \and G.A.~Mamon \at Institut d’Astrophysique de Paris (UMR 7095: CNRS \& Sorbonne Université), Paris, France  %modif name+affil
  \and B.~McArthur \at University of  Texas, Austin, TX, USA
  \and A.~Mora \at Aurora Technology BV, Madrid, Spain 
  \and M.~Shao \at Jet Propulsion Laboratory, California Institute of  Technology, Pasadena, CA, USA            % modif affil
  \and A.~Sozzetti \at Obs.\ Torino/INAF, Pino Torinese, Italy
  \and D.~Spolyar \at Stockholm University, Sweden
  \and E.~Villaver \at Centro de Astrobiología (CAB, C SIC-INTA), ESAC Campus, s/n/, 28692 Villanueva de la Cañada, Madrid, Spain
  \and U.~Abbas \at INAF/Obs. Torino, Pino Torinese, Italy
  \and C.~Albertus \at Universidad de Granada, Spain 
  \and J.~Alves \at University of Vienna, Vienna, Austria
  \and R.~Barnes \at Astronomy Department, University of Washington, Seattle, WA, USA                              % modif affil
  \and A.S.~Bonomo \at INAF, Obs.\ Torino, Pino Torinese, Italy 
  \and H.~Bouy \at Laboratoire d’Astrophysique de Bordeaux, Univ. Bordeaux, CNRS, Pessac, France     % modif affil
  \and W.R.~Brown \at CfA $|$ Harvard \& Smithsonian, Cambridge, MA, USA                                     % modif name + affil
  \and V.~Cardoso \at CENTRA, IST, Universidade de Lisboa, Lisbon, Portugal 
  \and M.~Castellani \at INAF - Rome Astronomical Obs., Roma, Italy 
  \and L.~Chemin \at University of Antofogasta, Antofogasta, Chile
  \and H.~Clark \at Univ. Sydney, Sydney, Australia
  \and A.C.M.~Correia \at CFisUC, Dept. Physics, University of Coimbra, Coimbra, Portugal                    %modif name + affil
  \and M.~Crosta \at Obs.\ Torino, INAF, Pino Torinese, Italy  
  \and A.~Crouzier \at LESIA, Observatoire de Paris, Université PSL, Sorbonne Université, Université de Paris, CNRS, Meudon, France
  \and M.~Damasso \at INAF, Obs.\ Torino, Pino Torinese, Italy 
  \and J.~Darling \at University of  Colorado, Boulder, CO, USA
  \and M.B.~Davies \at Lund University, Lund, Sweden 
  \and A.~Diaferio \at Dipartimento di Fisica, Università di Torino, Torino, Italy                                      % modif affil
                         \at Istituto Nazionale di Fisica Nucleare (INFN), Sezione di Torino, Torino, Italy
  \and M.~Fortin \at Copernicus Astronomical Center, Warsaw, Poland
  \and M.~Fridlund \at Leiden Obs., Leiden, The Netherlands 
                              \at Chalmers Univ., Gothenburg, Sweden
  \and M.~Gai  \at Obs.\ Torino, INAF, Pino Torinese, Italy  
  % \and A.~Gallo \at Dipartimento di Fisica, Università di Torino, Torino, Italy                                      % added + modif affil
  %                        \at Istituto Nazionale di Fisica Nucleare (INFN), Sezione di Torino, Torino, Italy
  \and P.~Garcia \at CENTRA, FEUP, Univ. Porto, Porto, Portugal                                                           % modif affil 
  \and O.~Gnedin \at University of  Michigan, Ann Arbor, MI, USA
  \and A.~Goobar \at Stockholm Univ., Sweden 
  \and P.~Gordo \at University of Lisbon - CENTRA/SIM, Lisbon, Portugal 
  \and R.~Goullioud \at Jet Propulsion Laboratory, California Institute of Technology, Pasadena, CA, USA        % modif affil
  \and D.~Hall \at The Open University, UK 
  \and N.~Hambly \at University of Edinburgh, UK 
  \and D.~Harrison \at IoA, Cambridge, UK 
  \and D.~Hobbs \at Lund Observatory, Department of Astronomy and Theoretical Physics, Lund University, Lund, Sweden  %modif affil
  \and A.~Holland \at The Open University, UK 
  \and E.~H\o{}g \at Niels Bohr Institute, Copenhague, Denmark 
  \and C.~Jordi \at Dept. FQA, Institut de Ciències del Cosmos (ICCUB), Universitat de Barcelona (IEEC-UB), Martí Franquès 1, E08028 Barcelona, Spain  % modif affil
  \and S.~Klioner \at Lohrmann Obs., Technische Universität Dresden, Dresden, Germany                    % modif affil
  \and A.~Lan\c{c}on \at University of Strasbourg, CNRS, Obs.\ Astron.\ Strasbourg, Strasbourg,  France 
  \and J.~Laskar \at IMCCE, Observatoire de Paris, Université PSL, Sorbonne Université, CNRS, Paris, France
  \and M.~Lattanzi \at Obs.\ Torino, INAF, Pino Torinese, Italy  
  \and C.~Le~Poncin-Lafitte \at SYRTE, Observatoire de Paris, Université PSL, CNRS, Sorbonne Université, LNE, 61 avenue de l’Observatoire 75014 Paris, France
  \and X.~Luri \at Dept. FQA, Institut de Ciències del Cosmos (ICCUB), Universitat de Barcelona (IEEC-UB), Martí Franquès 1, E08028 Barcelona, Spain  % modif affil
  \and D.~Michalik \at ESA/ESTEC, Noordwijk, The Netherlands
  \and A.~Moitinho~de~Almeida\at CENTRA, Faculdade de Ciências, Universidade de Lisboa, Lisbon, Portugal
  \and A.~Mour\~{a}o \at CENTRA, IST, University of Lisbon, Lisbon, Portugal 
  \and L.~Moustakas \at Jet Propulsion Laboratory, California Institute of Technology, Pasadena, CA, USA      %modif affil
  \and N.J.~Murray \at The Open University, UK                                                                                          % modif name (?)
  \and M.~Muterspaugh \at Dean of Science, Technology, and Mathematics,                                             % modif affil
                                            Columbia State Community College, Columbia, TN, USA 
  \and M.~Oertel \at LUTH, Observatoire de Paris, Université PSL, CNRS, Université de Paris, Meudon, France
  \and L.~Ostorero \at Dipartimento di Fisica, Università di Torino, Torino, Italy                                      % modif affil
                         \at Istituto Nazionale di Fisica Nucleare (INFN), Sezione di Torino, Torino, Italy
  \and J.~Portell \at Dept. FQA, Institut de Ciències del Cosmos (ICCUB), Universitat de Barcelona (IEEC-UB), Martí Franquès 1, E08028 Barcelona, Spain   % modif affil 
  \and J.-P.~Prost \at Thales Alenia Space, Cannes, France
  \and A.~Quirrenbach \at University of Heidelberg, Heidelberg, Germany 
  % \and K.A.~Rybicki \at Astronomical Observatory, University of Warsaw, Poland                                       % added+modif affil
  \and J.~Schneider \at LUTH, Observatoire de Paris, Université PSL, CNRS, Université de Paris, Meudon, France                                      % modif affil
  \and P.~Scott  \at University of  Queensland, Brisbane, Australia
                         \at Imperial College London, London, UK
  \and A.~Siebert \at University of Strasbourg, CNRS, Obs.\ Astron.\ Strasbourg,
  Strasbourg, France 
  \and A.~da~Silva \at Instítuto de Astrofísica e Ciências do Espaço, Faculdade de Ciências, Universidade de Lisboa, Lisbon, Portugal
  \and M.~Silva \at CENTRA/SIM-FEUP, University of Porto, Porto, Portugal 
  \and P.~Th\'ebault \at LESIA, Observatoire de Paris, Université PSL, Sorbonne Université, Université de Paris, CNRS, Meudon, France
  \and J.~Tomsick \at SSL Berkeley, Berkeley, CA, USA
  \and W.~Traub \at Jet Propulsion Laboratory, California Institute of Technology, Pasadena, CA, USA     % modif affil (+ deceased ?)
  \and M.~de Val-Borro \at Planetary Science Institute, Tucson, AZ, USA                                              % modif affil
  \and M.~Valluri \at University of Michigan, Ann Arbor, MI, USA
  \and N.A.~Walton \at Institute of Astronomy, University of Cambridge, UK                                      % modif affil 
  \and L.L.~Watkins \at AURA for the European Space Agency, ESA Office, Space Telescope Science Institute, 3700 San Martin Drive, Baltimore MD 21218, USA
  \and G.~White \at RAL Space, STFC Rutherford Appleton Laboratory, Chilton, Didcot, Oxfordshire, OX11 0QX, UK      % modif affil R2
                          \at Department of Physics and Astronomy, The Open University, Walton Hall, Milton Keynes, MK7 6AA, UK
  \and L.~Wyrzykowski \at Astronomical Observatory, University of Warsaw, Poland                          % modif affil
  \and R.~Wyse \at Johns Hopkins Univ., Baltimore, MD, USA
 \and Y.~Yamada \at Department of Physics, Kyoto University, Kyoto, Japan
}%

\date{Received: date / Accepted: date}

\maketitle

\begin{abstract}
  Sky survey telescopes and powerful targeted telescopes play
  complementary roles in astronomy. In order to investigate the nature
  and characteristics of the motions of very faint objects, a
  flexibly-pointed instrument capable of high astrometric accuracy is
  an ideal complement to current astrometric surveys and a unique tool
  for precision astrophysics. Such a space-based mission will push the
  frontier of precision astrometry from evidence of Earth-mass
  habitable worlds around the nearest stars, to distant
  Milky Way objects, and out to the Local Group of galaxies. As we
  enter the era of the James Webb Space Telescope and the new
  ground-based, adaptive-optics-enabled giant telescopes, by obtaining
  these high precision measurements on key objects that Gaia could not
  reach, a mission that focuses on high precision astrometry science
  can consolidate our theoretical understanding of the local Universe,
  enable extrapolation of physical processes to remote redshifts, and
  derive a much more consistent picture of cosmological evolution and
  the likely fate of our cosmos. Already several missions have been
  proposed to address the science case of faint objects in motion
  using high precision astrometry missions: NEAT proposed for the ESA
  M3 opportunity, micro-NEAT for the S1 opportunity, and Theia for the
  M4 and M5 opportunities. Additional new mission configurations
  adapted with technological innovations could be envisioned to pursue
  accurate measurements of these extremely small motions. The goal of
  this White Paper is to address the fundamental science questions
  that are at stake when we focus on the motions of faint sky objects
  and to briefly review instrumentation and mission profiles.
% \keywords: astrometry;  cosmology; local universe;  exoplanets; space mission
\end{abstract}

This paper is the accurate transcription with a very few updates of
the White Paper called \emph{Faint objects in motion: the new frontier
  of high precision astrometry} submitted in 2019 to the call for the
next planning cycle of the ESA Science Programme, called ``Voyage
2050''. Following ESA instructions, the main aim of this paper is to
argue why high precision astrometry should have priority in the Voyage
2050 planning cycle. In order to ensure realism in the resulting
Programme, we were asked to illustrate possible mission profiles which
is possible thanks to the previous work on the proposed space
missions.  Therefore, this paper focusses on the scientific issues
where most figures refer to the Theia specifications \citep[see][for
details]{Boehm2017} which target astrometric end-of-mission
precisions of 10\,$\mu$as for a faint object of $R=20$\,mag and
0.15\,$\mu$as for a bright object of $R=5$\,mag (see Table
\ref{tab:tech.summary.science}).

%%%%%%%%%%%%%%%%%%%%%%%%%%%%%%%%%%%%%%%%%%%%%%%%%%%%%%%%%%%%%%%%%%%%%
%%%%%%%%%%%%%%%%%%%%%%%%%%%%%%%%%%%%%%%%%%%%%%%%%%%%%%%%%%%%%%%%%%%%%
\section{Science questions}
\label{sec:science-questions}

Europe has long been a pioneer of astrometry, from the time of ancient
Greece to Tycho Brahe, Johannes Kepler, the Copernican revolution and
Friedrich Bessel. ESA's Hipparcos \citep{1997A&A...323L..49P,
  2000A&A...355L..27H} and Gaia \citep{2016A&A...595A...1G} satellites
continued this tradition, revolutionizing our view of the Solar
Neighborhood and Milky Way, and providing a crucial foundation for
many disciplines of astronomy.

An unprecedented microarcsecond relative precision mission will
advance European astrometry still further, setting the stage for
breakthroughs on the most critical questions of cosmology, astronomy,
and particle physics.

%%%%%%%%%%%%%%%%%%%%%%%%%%%%%%%%%%%%%%%%%%%%%%%%%%%%%%%%%%%%%%%%%%%%%
\subsection{Dark matter}
\label{sec:dark-matter}

The current hypothesis of cold dark matter (CDM) urgently
  needs verification. Dark matter (DM) is essential to the $\Lambda$
+ CDM cosmological model ($\Lambda$CDM), which successfully describes
the large-scale distribution of galaxies and the angular fluctuations
of the Cosmic Microwave Background, as confirmed by ESA's Planck
mission \citep{2011A&A...536A...1P}. Dark matter is the dominant form
of matter ($\sim 85\%$) in the Universe, and ensures the formation and
stability of enmeshed galaxies and clusters of galaxies. The current
paradigm is that dark matter is made of heavy, hence cold, particles,
otherwise galaxies will not form. However, the nature of dark
  matter is still unknown.

There are a number of open issues regarding $\Lambda$CDM on
small scales. Simulations based on DM-only predict 1) a large number
of small objects orbiting the Milky Way, 2) a steep DM distribution in
their centre and 3) a prolate Milky Way halo. However, hydrodynamical
simulations, which include dissipative gas and powerful astrophysical
phenomena (such as supernovae explosions and jets from galactic
nuclei) can change this picture. Quantitative predictions are based on
very poorly understood sub-grid physics and there is no consensus yet
on the results. Answers are buried at small-scales, which are
extremely difficult to probe. A new high precision astrometric mission
appears to be the best way to settle the nature of DM and will allow
us to validate or refute key predictions of $\Lambda$CDM, such as
\begin{itemize}
\item the DM distribution in dwarf spheroidal galaxies  
\item the outer shape of the Milky Way DM halo 
\item the lowest masses of the Milky Way satellites and subhalos
\item the power spectrum of density perturbations
\end{itemize}
These observations will significantly advance research in\-to DM. They
may indicate that DM is warmer than $\Lambda$CDM predicts. Or we may
find that DM is prone to self-interac\-tions that reduces its density
in the central part of the satellites of the Milky Way. We may
discover that DM has small interactions that reduce the number of
satellite companions. Alternatively, measurement of the Milky Way DM
halo could reveal that DM is a sophisticated manifestation of a
modification of Einstein's gravity.

\subsubsection{ The DM distribution in dwarf spheroidal galaxies}

\begin{figure}[t]
  \centering
  \includegraphics[width=0.8\hsize]{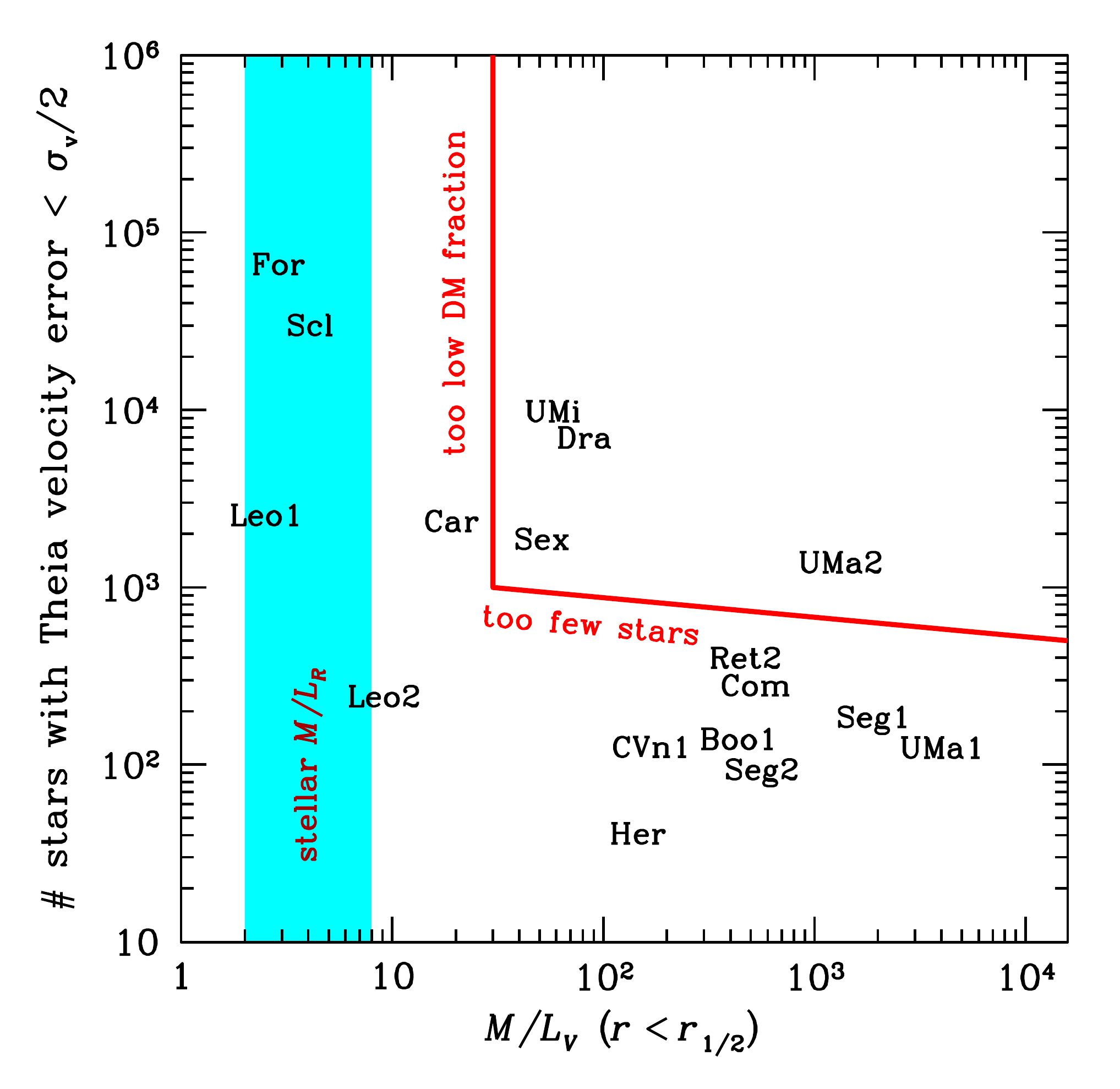}
  \caption{Number of dwarf spheroidal galaxy stars within the
    field of view of Theia, a high precision astrometry concept, with expected
    plane-of-sky errors lower than half the galaxy's velocity
    dispersion as a function of the galaxy's estimated mass-to-light
    ratio within the effective (half-projected-light) radius of the
    galaxy.  Luminosities and total masses within the half-light radii
    are mainly from \citet{Walker+09}.}
  \label{fig:NvsMoverL}
\end{figure}

\begin{figure*}[t]
\centering
\includegraphics[width=0.45\hsize,clip]{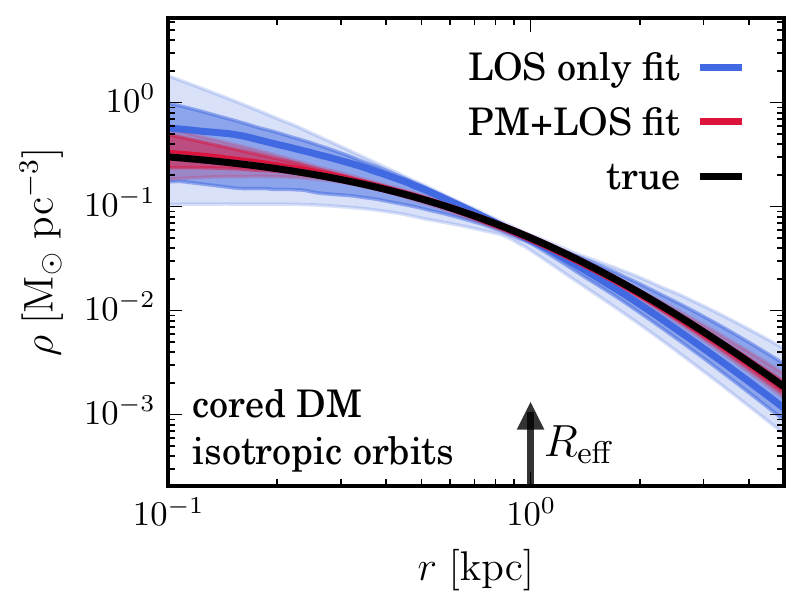}
\includegraphics[width=0.45\hsize,clip]{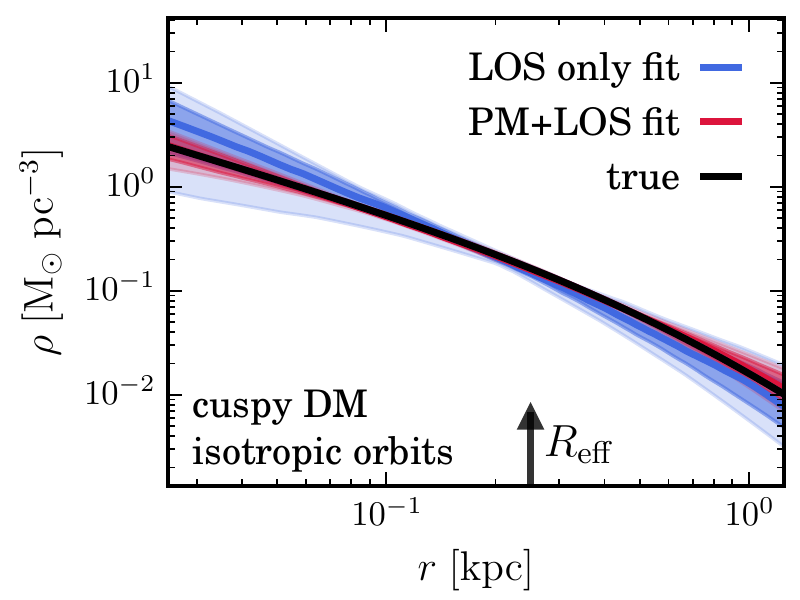}
\includegraphics[width=0.45\hsize,clip]{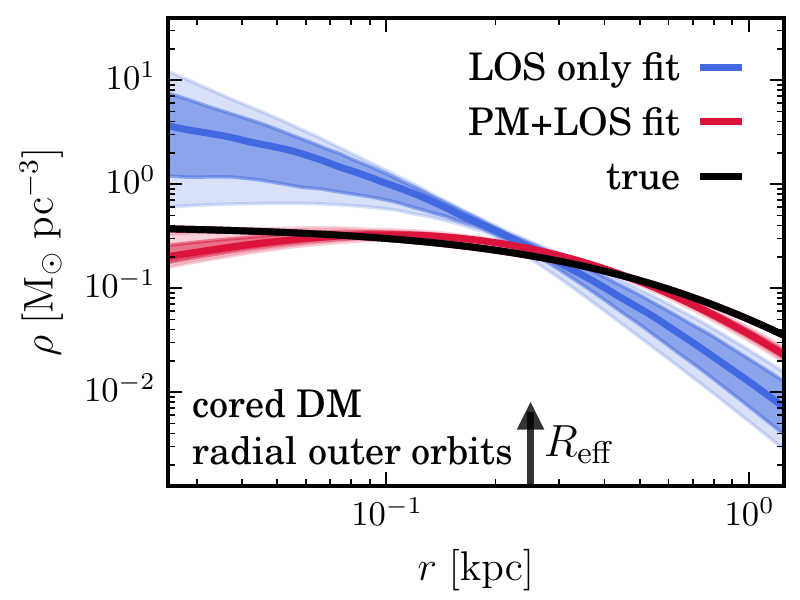}
\includegraphics[width=0.45\hsize,clip]{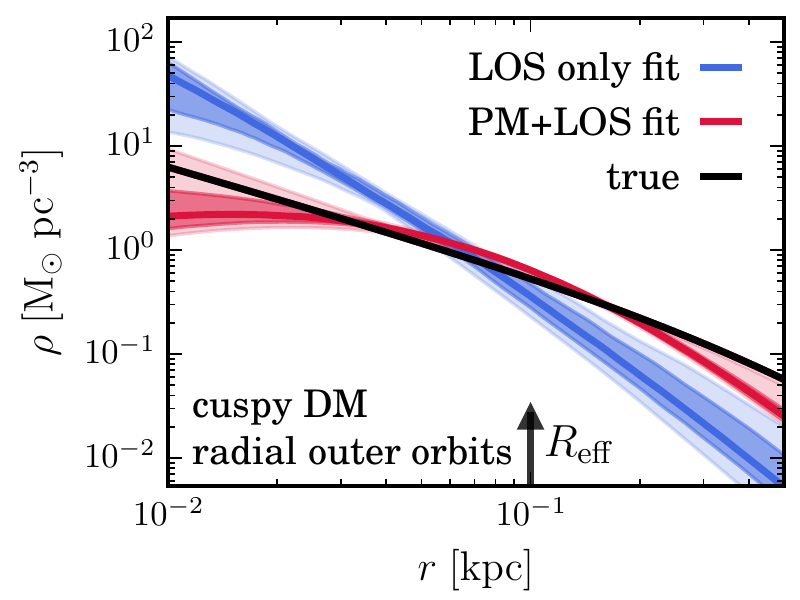}
 \caption{Reconstruction of the DM halo profile of the Draco dSph
  without (\textit{blue}) and with (\textit{red}) proper motions using
  the mass-orbit modeling algorithm of \citet{2013MNRAS.436.2598W}. Four
  mocks of Draco \citep{2011ApJ...742...20W} were used, with cored
  (\textit{left}) and cuspy (\textit{right}) DM halos, and with
  isotropic velocities everywhere (\textit{top}) or only in the inner
  regions with increasingly radial motions in the outer regions
  (\textit{bottom}). The effective (half-projected light) radii of
  each mock is shown with the \textit{arrows}. The stellar proper
  motions in the mocks were perturbed with apparent magnitude
  dependent errors as expected with 1000 hours of observations spread
  over 4 years.}
\label{fig:rhoofrPM}
\end{figure*}

Because they are DM-dominated (see Fig.~\ref{fig:NvsMoverL} where the
number of stars versus the mass-to-light ratio is present\-ed), dwarf
Spheroidal galaxies (dSphs) are excellent laboratories to test the
distribution of DM within the central part of small galaxies and
disentangle the influence of complex baryonic processes from that of
DM at these scales.

Simulations from \citet{Onorbe+15} or \citet{Read+16} for example show
that the DM distribution (referred to as DM profile) in dSphs strongly
depends on their star formation history. More specifically, these
simulations find that CDM can be heated by bursty star formation
inside the stellar half-light radius $r_{1/2}$, if star formation
proceeds for long enough. As a result, some dSphs like Fornax have
formed stars for almost a Hubble time and so should have large central
DM cores, while others, like Draco and Ursa Major 2, had their star
formation truncated after just $\sim 1-2$\,Gyrs and should retain
their steep central DM cusp.

Large DM cores could also be attributed however to strong
self-interactions. Hence finding evidence for such cores in the
faintest dSphs (which are even more DM dominated \citep{Wolf+10} than
the classical ones), will bring tremendous insights about the history
of baryonic process\-es in these objects and could even dramatically
change our understanding of the nature of DM. Indeed,
self-inter\-acting DM \citep{Spergel&Steinhardt00} is expected to
scatter in the dense inner regions of dSphs, and thus leads to
homogeneous cores. Finding such a core DM distribution in dSphs could
then reveal a new type of particle forces in the DM sector and provide
us with new directions to build extensions of the Standard Model of
particle physics. On the other hand, finding cuspy DM profiles in all
dSphs (including the faintest ones) will confirm $\Lambda$CDM and
place strong constraints on galaxy formation. As shown in
Figs.~\ref{fig:vposerrvsDist} and \ref{fig:astroComparisons}, a
telescope with microarcsecond astrometric precision allows us to
determine whether the DM distribution in dSphs is cuspy or has a core,
and hence can lead to a very significant breakthrough regarding the
nature of DM.
 
To determine the inner DM distribution in dSphs, one needs to remove
the degeneracy between the radial DM profile and orbital anisotropy
that quantifies whether stellar orbits are more radial or more
tangential in the Jeans equation \citep{Binney&Mamon82}.  This can be
done by adding the proper motions of stars in dSphs.
Fig.~\ref{fig:rhoofrPM} shows that for the Draco dSph (which was
obtained using single-component spherical mock datasets from the Gaia
Challenge Spherical and Triaxial Systems working group,\footnote{See
  \url{http://astrowiki.ph.surrey.ac.uk/dokuwiki/doku.php?id=tests:sphtri}}
and the number of stars expected to be observed by a high precision
astrometry mission), the inclusion of proper motions lifts the cusp /
core degeneracy that line-of-sight-only kinematics cannot disentangle.

We remark in addition that a high precision astrometric mission is
able to perform follow-ups of Gaia's observations of dSphs streams of
stars if needed. Not only will such a mission provide the missing
tangential velocities for stars with existing radial velocities, but
it will also provide crucial membership information - and tangential
velocities - for stars in the outer regions of the satellite galaxies
that are tidally disrupted by the Milky Way.

%%%%%%% Oleg, Warren
\subsubsection{ The triaxiality of the Milky Way dark matter halo \label{triaxial} }

For almost two decades cosmological simulations have shown that Milky
Way-like DM halos have triaxial sha\-pes, with the degree of triaxiality
varying with radius \citep[][for example]{dubins_94, kkzanm04}: halos
are more round or oblate at the center, become triaxial at
intermediate radii, and prolate at large radii \citep{zemp_etal_11}.

Precise measurement of the velocity of distant Hyper Velocity Stars
(hereafter HVS) can test these departures from spherical symmetry,
independently of any other technique attempted so far (such as the
tidal streams). HVSs were first discovered serendipitously
\citep{Brown+05,Hirsch+05,Edelmann+05}, and later discovered in a
targeted survey of blue main-sequence stars \citep[][and references
there\-in]{Brown15_ARAA}. Gaia measurements demonstrate that candidate
hyper velocity stars include unbound disc runaways \citep{2019A&A...628L...5I},
unbound white dwarfs ejected from double-degenerate type Ia supernovae
\citep{Shen2018}, and runaways from the LMC \citep{Erkal2018}, however
the highest-velocity main sequence stars in the Milky Way halo have
trajectories that point from the Galactic Center \citep{Brown2018,
  2020MNRAS.491.2465K}.  \vspace{5mm}

\begin{figure}[thb]
\centering
\fbox{\includegraphics[width=0.9\hsize,clip]{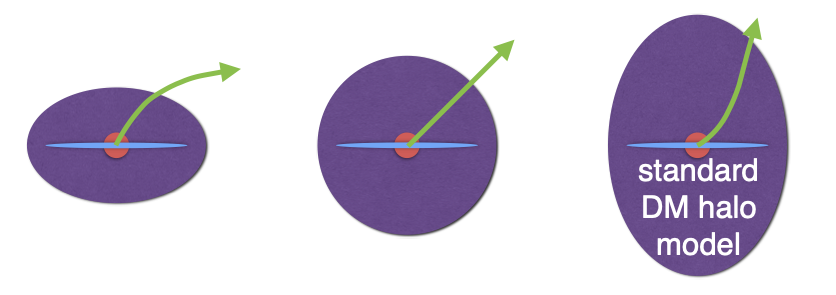}}
 \caption{Illustration of the trajectories of hyper velocity stars
  ejected from Galactic Center for 3 different outer DM halo shapes:
  oblate (\textit{left}), spherical (\textit{middle}), and prolate
  (\textit{right}).}
\label{Fig:triaxial_halos}
\end{figure}

Because these velocities exceed the plausible limit for a runaway star
ejected from a binary, in which one component has undergone a
supernova explosion, the primary mechanism for a star to obtain such
an extreme velocity is assumed to be a three-body interaction and
ejection from the deep potential well of the supermassive black hole
at the Galactic Center \citep{Hills88,Yu&Tremaine03}.

\vspace{8mm}
\begin{figure}[bht]
  \centering
  \includegraphics[width=0.8\hsize]{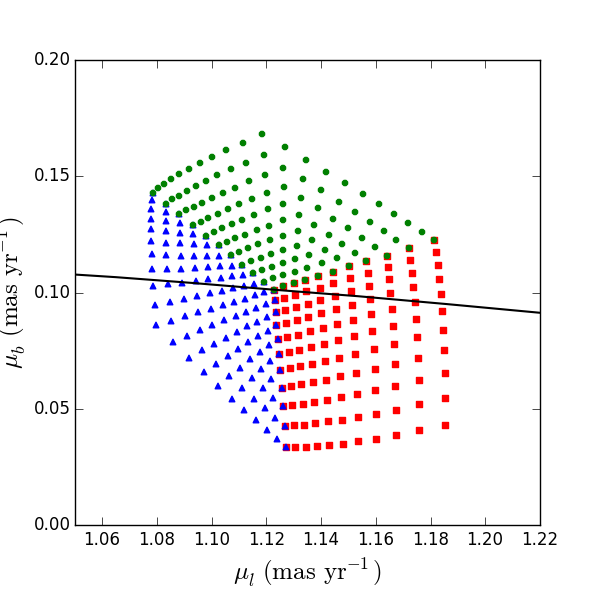}
  \caption{Expected proper motions of HVS5 under different
    assumptions about the shape and orientation of the DM halo. The
    families of models are shown with the halo major axis along the
    Galactic X- (\textit{red squares}), Y- (\textit{blue triangles}),
    and Z- (\textit{green circles}) coordinates.  The \textit{solid
      line} shows how the centroid of the proper motions will shift
    with a different distance to HVS5.}
\label{fig:pm_hvs5} 
\end{figure}

By measuring the three-dimensional velocity of these stars, we will
reconstruct the triaxiality of the Galactic potential. In a spherical
potential, unbound HVS ejected from the Galactic Center should travel
in nearly a straight line, as depicted in
Fig.\ref{Fig:triaxial_halos}.  However, for triaxial halos, the
present velocity vector should not point exactly from the Galactic
Center because of the small curvature of the orbit caused by
non-spherically symmetric part of the potential
\citep{Gnedin+05,Yu&madau07}. While both the halo and stellar disc
induce transverse motions, the effect is dominated by halo triaxiality
at the typical distance of HVSs. The deflection contributed by the
disc peaks around 10 kpc but quickly declines at larger distances,
while the deflection due to the triaxial halo continues to accumulate
along the whole trajectory. Fig.~\ref{fig:pm_hvs5} shows the
spread of proper motion for one star, HVS5, for different halo shapes
(different halo axis ratios and different orientations of the major
axis).

Proper motions of several HVSs were measured with the Hubble Space
Telescope (HST) by \cite{Brown+15}, using an astrometric frame based
on background galaxies. However, these measurements were not
sufficiently accurate to constrain the halo shape or the origin of
each HVS.  A high precision astrometric mission with a sufficiently
large field of view could include about 10 known quasars from the SDSS
catalog around most HVSs. This will provide a much more stable and
accurate astrometric frame, and will allow us to constrain the halo
axis ratios to about 5$\%$.

\begin{figure}[bht]
\centering
\includegraphics[width=0.9\hsize, trim=0.cm 0.3cm 0.8cm 1cm, clip]{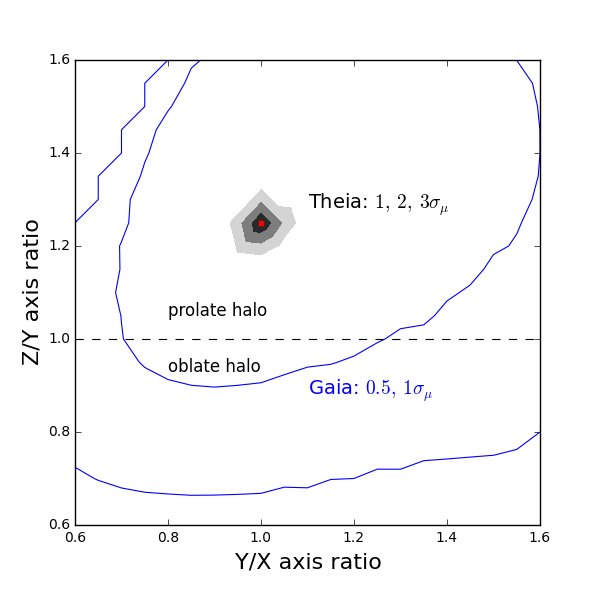}
 \caption{Example of a reconstruction of the Galactic halo shape
  from a high precision astrometry mission (Theia) measurement of
  proper motion of HVS5.  The assumed proper motions correspond to a
  prolate model with $q_X = q_Y = 0.8\, q_Z$, marked by a \textit{red
    square}. \textit{Shaded contours} represent confidence limits
  corresponding to the expected 1, 2, and $3\, \sigma_\mu$ proper
  motion errors. The \textit{outer blue contours} show the accuracy
  that will be achieved by Gaia at the end of its mission, even if its
  expected error was reduced by a factor of
  2.} \label{Fig:axishvs5d20theia}
\end{figure}

%%%%% ARNAUD's figure  -- moved to appear on right page
\begin{figure*}[t]
\centering
\includegraphics[width=0. \hsize]{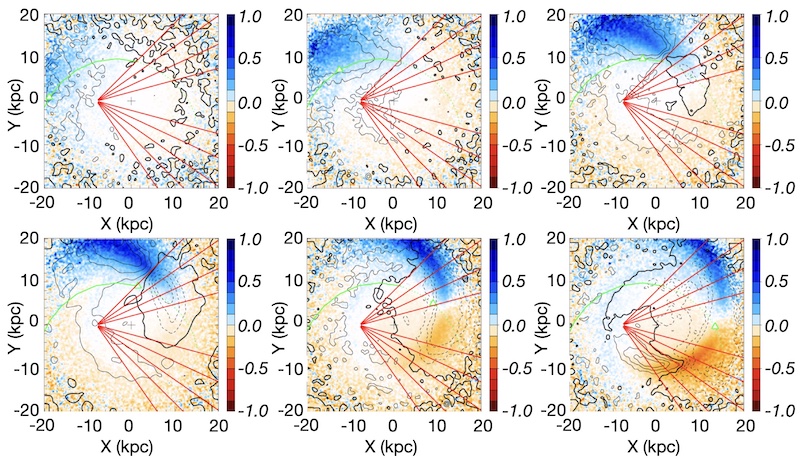}
 \caption{Face-on view of the evolution of the perturbation of a
  Galactic Disc due to a DM subhalo of mass 3$\%$ of the mass of the
  disc crossing the disc from above. The \textit{upper} and
  \textit{lower} panels are before and after the crossing,
  respectively, for different times 125, 75 and 25 Myr before the
  crossing and 25,75,125 Myr after (from \textit{left} to
  \textit{right}).  The mean displacement amplitude is indicated in
  the color bar, while the \textit{contours} indicate the amplitude of
  the bending mode in velocity space, using plain lines for positive
  values and dashed lines for negative values. The \textit{green line}
  shows the projected orbit of the subhalo (dashed line after the
  impact with the disc). The \textit{green triangle} shows the current
  location of the subhalo on its orbit.  The \textit{red lines} are
  our potential lines of sight for Theia, a high precision astrometry
  mission concept, spaced by 10$^\circ$ in longitude with one pointing
  above the plane and one below the plane, that will allow us to map
  the disc perturbation behind the Galactic Center. }
\label{fig:dm:1}
\end{figure*}
%%%%%% Arnaud  %edit barbara
% we need a standard for subsection labels capitalized or not
% section had disc spelled different ways and nouns capitalized
% variably.   expanded captio n to include some body text do not repeat

Fig.\ref{Fig:axishvs5d20theia} shows that with a precision of
$4 \, \mu$as/yr one can constrain the orientation of the halo major
axis and measure the axis ratios to an accuracy of
$\delta (q_Z/q_X) < 0.05$ for the typical HVS distance of 50 kpc.  For
comparison, Gaia at the end of its mission will achieve only
$40 - 150 \, \mu$as/yr, which is insufficient to provide useful
constraints on the axis ratios.

Statistical studies of high-precision proper motions of HVSs can also
constrain departures of the halo shape from spherical \citep[][ in
preparation]{Gallo2021}. Indeed, numerical simulations of the
trajectories of synthetic HVSs ejected through the Hills mechanism
show that the distributions of the HVS tangential velocities in the
Galactocentric reference frame are significantly different from
spherical and non-spherical halos: the significance is
$P \le 1.3 \times 10^{-6}$ for oblate halos with $q_Z/q_X \le 0.9$ and
$P \le 2.2 \times 10^{-5}$ for prolate halos with $q_Z/q_X \ge
1.1$. The median tangential velocity of a sample of $\sim 100$ HVSs
located at heliocentric distances $\sim 50\,\mbox{kpc}$ can differ by
$\sim 5-10\,\mbox{km/s}$, implying differences in proper motions of
$\sim 20-40\,\mbox{$\mu$as/yr}$ between spherical and non-spherical
halos.

Finally, an accurate measurement of HVS velocities may lead to
improved understanding of the black hole(s) at the Galactic
Center. Indeed, theoretical models show that HVSs will have a
different spectrum of ejection velocities from a binary black hole
versus a single massive black hole. Gaia has led to the discovery of
several candidate hypervelocity stars \citep[ejection velocities of
over $550\,\mbox{km/s}$:][]{2018A&A...620A..48I, 2018ApJ...869...33H,
  2019ApJ...881..106M}, that were definitely not ejected from the
Galactic Center but were ejected from spiral arms in the Milky Way
disc. These possibly point to intermediate mass black holes of mass
$~100\,\mbox{M}_\odot$ - these could be local remnants of binary black
hole mergers of the kind discovered by LIGO/Virgo
\citep{2009RPPh...72g6901A, 2015CQGra..32b4001A} and could have
important implications for our understanding of stellar evolution.

%%%%%%% Monica
\subsubsection{Orbital distribution of Dark Matter from the orbits of halo stars}

The orbits of DM particles in halos\footnote{For an analysis of
  orbital content of DM halos see \cite{valluri_etal_10,
    valluri_etal_12, valluri_etal_13, bryan_etal_12}.} cannot be
detected directly since DM particles interact only weakly with normal
matter. However, in a triaxial potential such as described above, it
is expected that a large fraction of the DM orbits do not have any net
angular momentum. Hence these particles should get arbitrarily close
to the center of the cusp, regardless of how far from the center they
were originally. This allows DM particles, which annihilate within the
cusp to be replenished on a timescale $10^4$ times shorter than in a
spherical halo \citep[analogous to loss cone filling in the case of
binary black holes, see][]{merritt_poon_04}.

Recent work on the orbital properties and kinematic distributions of
halo stars and DM particles show that halo stars, especially the ones
with lowest metallicities, are relatively good tracers of DM particles
\citep{2013ApJ...767...93V, 2018PhRvL.120d1102H, 2018JCAP...04..052H}
and observations with Gaia Data Release 2 (DR2) may have already led
to the kinematic discovery of dark substructure \citep{Necib2019}.
The orbits reflect both the accretion/formation history and the
current shape of the potential because DM halos are dynamically young
(i.e. they are still growing and have not attained a long term
equilibrium configuration where all orbits are fully phase mixed).
This opens up the very exciting possibility that one can infer the
kinematical distribution of DM particles by assuming that they are
represented by the kinematics of halo stars.

\subsubsection{Perturbations by Dark Matter subhalos}

A central prediction of $\Lambda$CDM in contrast to many alternatives
of DM, such as warm DM \citep[e.g.][]{Schaeffer:1984bt} or interacting
DM \citep[e.g.][]{Boehm:2014vja}, is the existence of numerous $10^6$
to $10^8$ M$_\odot$ DM \emph{subhalos} in the Milky Way halo. Their
detection is extremely challenging, as they are very faint and lighter
than dSphs. However, N-body simulations of the Galactic Disc show that
such a DM halo passing through the Milky Way disc will warp the disc
and produce a motion (bending mode), as shown in
Fig.~\ref{fig:dm:1}. This opens new avenues for detection as such
perturbations of the disc will result in anomalous motions of the
stars in the disc \citep[e.g.][for recent
analysis]{2015MNRAS.446.1000F}, that could give rise to an astrometric
signal.

\begin{figure*}[ht]
  \centering
  \includegraphics[width=0.9\hsize]{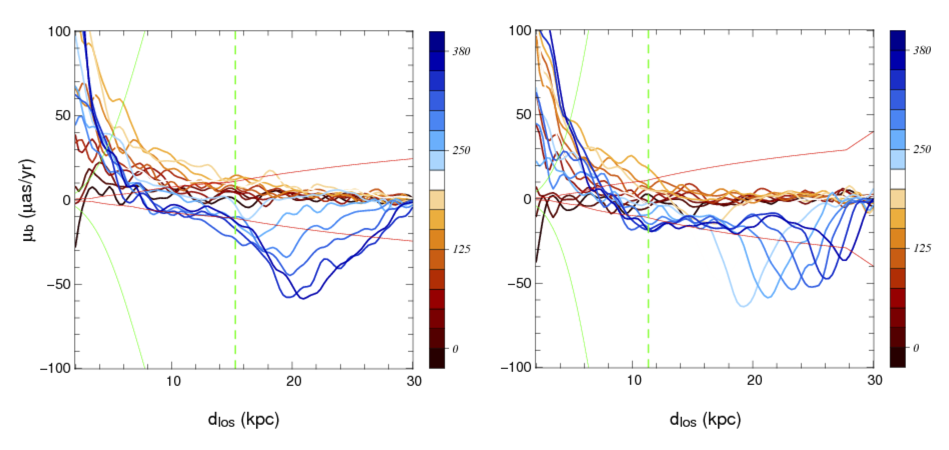}
  \caption{Astrometric signatures in the proper motion along
    Galactic latitude of the perturbation of disc stars by a
    subhalo. The \textit{left} and \textit{right} panels show lines of
    sight as a function of distance along the line of sight and time,
    for $\ell=-25^\circ$ and $\ell=+25^\circ$ respectively for
    $b=+2^\circ$. The color codes the time in Myr, \textit{red} for
    times prior to the crossing of the plane by the satellite,
    \textit{blue} for later times. The \textit{green line} is Gaia's
    expected end of mission performance for a population of red clump
    stars along these lines of sight. The \textit{vertical dashed
      line} is Gaia's detection limit ($G$=20) for the same
    population. The \textit{red lines} are Theia's expected 1$\sigma$
    accuracy for the same stars and for a 400~h exposure of the field
    over the course of the mission.  }
\label{fig:dm:2}
\end{figure*}

%% Arnaud's figure was moved up to appear on right page
These anomalous bulk motions develop both in the solar vicinity
\citep{2012ApJ...750L..41W} and on larger scales
\citep{2015MNRAS.446.1000F}, see Fig.\ref{fig:dm:2}. Therefore,
measuring very small proper motions of individual faint stars in
different directions towards the Galactic Disc could prove the
existence of these subhalos and confirm the CDM
scenario. Alternatively, in case they are not found, high precision
astrometric observations will provide tantalizing evidence for
alternative DM scenarios, the most popular today being a warmer form
of DM particle, though these results could also indicate DM
interactions \citep{Boehm:2014vja}.

A field of view of $1^\circ\times 1^\circ$ in the direction of the
Galactic Disc has $\sim 10^6$ stars with an apparent magnitude of
$R \leq 20$ (given by the confusion limit). Given the astrometric
precisions per field of view of Fig.~\ref{fig:astroComparisons}, a
high precision astrometric instrument could detect up to 7 impacts on
the disc from sub-halos as small as a few $10^6\,\mbox{M}_\odot$.

Gaia DR2 astrometry has led to the discovery of gaps in tidal streams
\citep{2018ApJ...863L..20P} like the GD1 stream. The gaps and
off-stream stars (spur) are consistent with gravitational interactions
with compact DM subhalos. Furthermore, Gaia DR2 data have revealed
that globular cluster streams (GD1 and Jhelum) show evidence for
cocoon-like structures that most likely arise from evolution inside a
(dark) subhalo prior to their tidal disruption by the Milky Way itself
\citep{2020ApJ...889..107C, 2019ApJ...881..106M, 2019ApJ...881L..37B}.
The high astrometric precision of a {Theia}-like mission will enable
us to measure the small velocity perturbations around the gaps in
streams and allow for a much more accurate determination of both the
masses and density structures of the perturbing dark subhalos.

%%%%%%% Pat, Hamish, Adrienne
\subsubsection{ Ultra-compact minihalos of dark matter in the Milky Way} 
\label{sec:science_minihalos}

In the $\Lambda$CDM model, galaxies and other large-scale structures
formed from tiny fluctuations in the distribution of matter in the
early Universe. Inflation predicts a spectrum of primordial
fluctuations in the curvature of spacetime, which directly leads to
the power spectrum of initial density fluctuations.  This spectrum is
observed on large scales in the cosmic microwave background and the
large scale structure of galaxies, but is very poorly constrained on
scales smaller than 2\,Mpc.  This severely restricts our ability to
probe the physics of the early Universe.  A high precision astrometric
mission could provide a new window on these small scales by searching
for astrometric microlensing events caused by \emph{ultra-compact
  minihalos} (UCMHs) of DM.

UCMHs form shortly after matter domination (at $z\sim1000$), in
regions that are initially overdense \citep[e.g.
$\delta\rho/\rho > 0.001$ in][]{2009ApJ...707..979R}.  UCMHs only form
from fluctuations about a factor of 100 larger than their regular
cosmological counterparts, so their discovery will indicate that the
primordial power spectrum is not scale invariant.  This will rule out
the single-field models of inflation that have dominated the
theoretical landscape for the past thirty years.  Conversely, the
absence of UCMHs can be used to establish upper bounds on the
amplitude of the primordial power spectrum on small scales
\citep{Bringmann11}, which will rule out inflationary models that
predict enhanced small-scale structure \citep{2016PhRvL.117n1102A}.

\begin{figure}[htb]
  \centering
  \includegraphics[width=0.8\hsize]{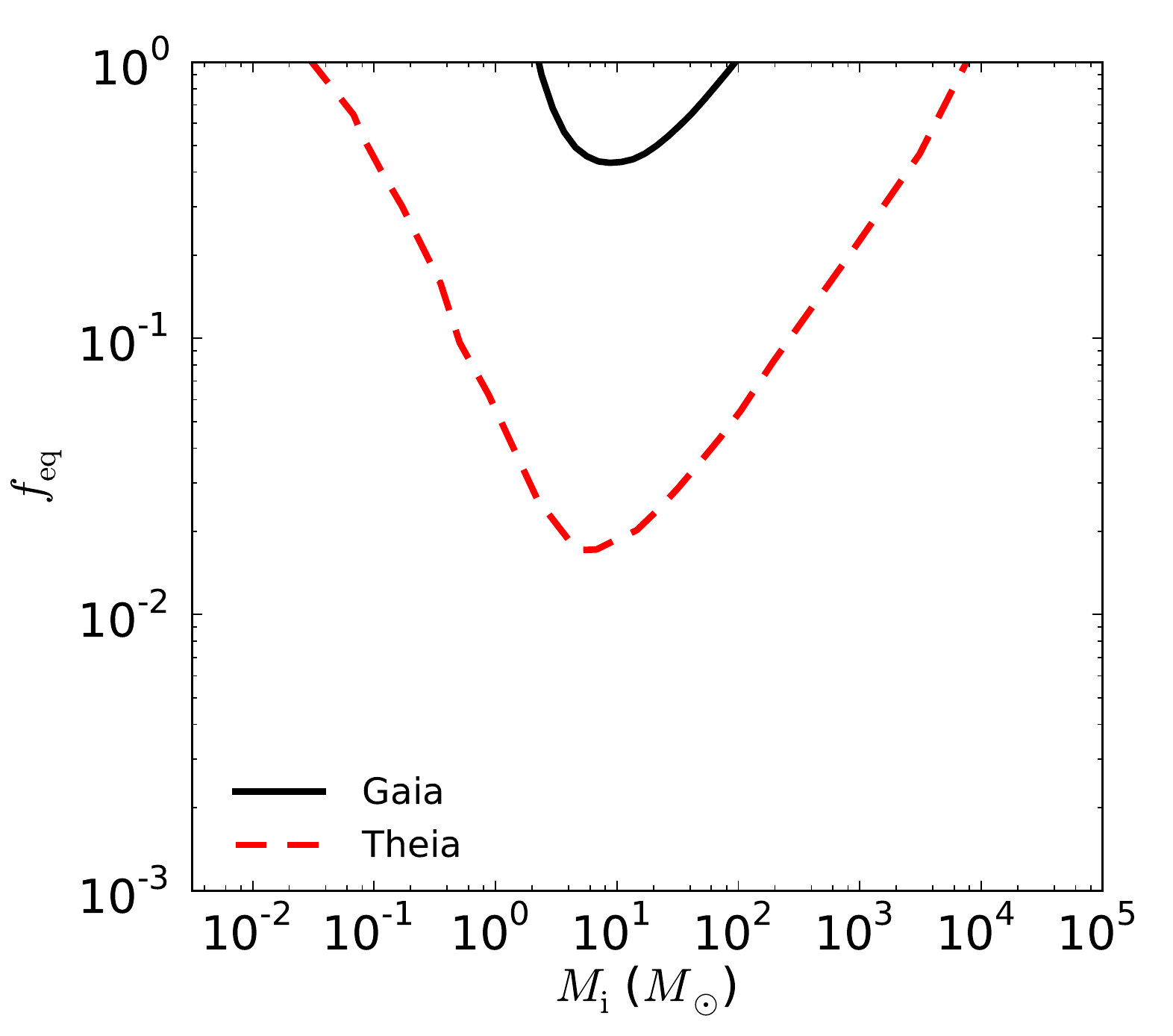}
  \caption{Projected sensitivity of a high precision astrometry
    mission (Theia) to the fraction of dark matter in the form of
    ultra-compact minihalos (UCMHs) of mass $M_i$ at the time of
    matter-radiation equality.  Smaller masses probe smaller scales,
    which correspond to earlier formation times (and therefore to
    \textit{later} stages of inflation). A UCMH mass of 0.1\,M$_\odot$
    corresponds to a scale of just 700\,pc.  Expected constraints from
    Gaia are given for comparison, showing that a {Theia}-like mission
    will provide much stronger sensitivity, as well as probe smaller
    scales and earlier formation times than ever reached before.}
  \label{fig:ucmh}
\end{figure}

\begin{figure*}[t]
    \centering
  \includegraphics[width=0.8\hsize,trim = 0cm 6.3cm 0cm 7.5cm,clip]{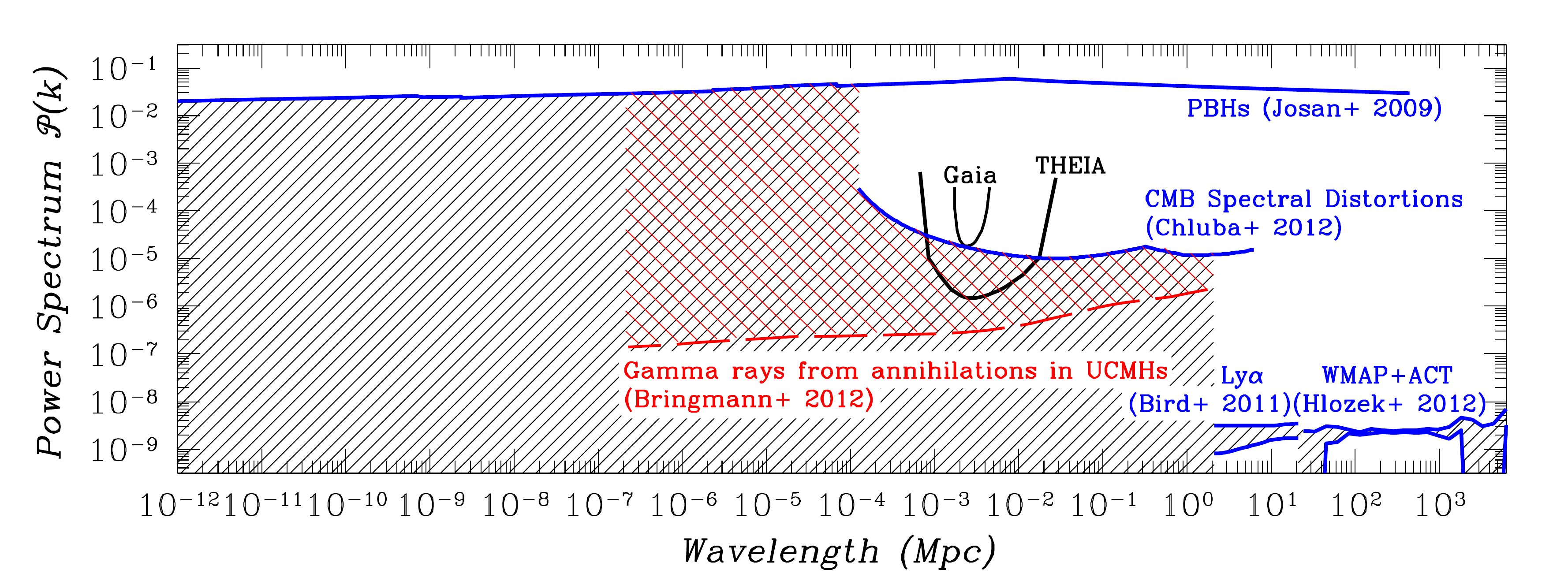}
  \caption{Limits on the power of primordial cosmological
    perturbations at all scales, from a range of different sources. A
    {Theia}-like mission will provide far stronger sensitivity to
    primordial fluctuations on small scales than Gaia, spectral
    distortions or primordial black holes (PBHs).  Unlike gamma-ray
    UCMH limits, a high precision astrometry mission's sensitivity to
    cosmological perturbations will also be independent of the
    specific particle nature of dark matter.}
  \label{fig:ucmh2}
\end{figure*}

Like standard DM halos, UCMHs are too diffuse to be detected by
regular photometric microlensing sear\-ches for MAssive Compact Halo
Objects (MACHOs).  Because they are far more compact than standard DM
halos, they however produce much stronger \textit{astrometric}
microlensing signatures \citep{2012PhRvD..86d3519L}.  By searching for
microlensing events due to UCMHs in the Milky Way, a high precision
astrometric mission will provide a new probe of inflation.  A search
for astrometric signatures of UCMHs in the Gaia dataset could
constrain the amplitude of the primordial power spectrum to be less
than about $10^{-5}$ on scales around 2\,kpc
\citep{2012PhRvD..86d3519L}.  Fig.\ \ref{fig:ucmh} shows that higher
astrometric precision (corresponding to that of
Fig.~\ref{fig:astroComparisons}) will provide more than \textit{an
  order of magnitude higher sensitivity} to UCMHs, and around
\textit{four orders of magnitude greater mass coverage} than Gaia.
These projections are based on 8000\,hr of observations of 10 fields
in the Milky Way disc, observed three times a year, assuming that the
first year of data is reserved for calibrating stellar proper motions
against which to look for lensing perturbations.  Fig.\
\ref{fig:ucmh2} shows that a high precision astrometric mission will
test the primordial spectrum of perturbations down to scales as small
as 700\,pc, and improve on the expected limits from Gaia by over an
order of magnitude at larger scales.

The results will be independent of the DM nature, as astrometric
microlensing depends on gravity only, unlike other constraints at
similar scales based on DM annihilation, from the Fermi Gamma-Ray
Space Telescope \citep{Bringmann11}.  An astrometric mission with
higher precision (shown in Fig.~\ref{fig:astroComparisons}) will have
sensitivity four orders of magnitude better than constraints from the
absence of primordial black holes
\citep[PBHs;][]{2009PhRvD..79j3520J}, and more than an order of
magnitude better than CMB spectral distortions
\citep{2012ApJ...758...76C}, which give the current best
model-independent limit on the primordial power spectrum at similar
scales.

% These are needed because they are cited in Fig \ref{fig:ucmh2}
\nocite{2011MNRAS.413.1717B, 2012ApJ...749...90H, 2009PhRvD..79j3520J}

\subsubsection{Directly Testing Gravity} 

Using the nearest star, Proxima Centauri, astrometry could measure the
behaviour of gravity at low accelerations.  A high precision
astrometry mission with an extended baseline of 10 years and a
precision of 0.5 $\mu$as could measure the wide binary orbit of
Proxima Centauri around Alpha Centauri A and B to distinguish between
Newtonian gravity and Milgromian dynamics (MOND).  The separation
between Proxima Centauri and the Alpha Centauri system suggests orbital
acceleration that is significantly less than the MOND acceleration
constant $a_0 \sim 1.2 \times 10^{-10}$ m/s$^2$ \citep{Banik2019}.  It
would be the first direct measurement of the departure from Newtonian
gravity in the very weak field limit, as expected in MOND, and the
results could have profound implications for fundamental physics.

%%%%%%%%%%%%%%%%%%%%%%%%%%%%%%%%%%%%%%%%%%%%%%%%%%%%%%%%%%%%%%%%%%%%%
\subsection{Exoplanets}
\label{sec:exoplanets}

\subsubsection{The Frontier of Exoplanet Astrophysics}

The ultimate exoplanetary science goal is to answer the enigmatic and
ancient question, ``\emph{Are we alone?'}' via unambiguous detection
of biogenic gases and molecules in the atmosphere of an Earth twin
around a Sun-like star \citep{Schwieterman+16}. Directly addressing
this age-old question related to the uniqueness of the Earth as a
habitat for complex biology constitutes today the vanguard of the
field, and it is clearly recognized as an unprecedented,
cross-technique, interdisciplinary endeavor.

Since the discovery of the first Jupiter-mass companion to a
solar-type star \citep{Mayor&Queloz95}, tremendous progress has been
made in the field of exo\-planets. Our knowledge is expanding quickly
due to the discovery of thousands of planets, and the skillful
combination of high-sensitivity space-borne and ground-based programs
that have unveiled the variety of planetary systems architectures that
exist in the Galaxy
\citep[e.g.][]{2013Sci...340..572H,2011arXiv1109.2497M}.  Preliminary
estimates \citep[e.g.][]{Winn&Fabrycky15} are now also available for
the occurrence rate $\eta_\uplus$ of terrestrial-type planets in the
Habitable Zone (HZ) of stars more like the Sun ($\eta_\uplus\sim10\%$)
and low-mass M dwarfs ($\eta_\uplus\sim50\%$).

However, transiting or Doppler-detected HZ terrestrial planet
candidates \citep[including the discovery of the $m_{\rm p}\sin i=1.3$
$M_\oplus$ HZ-planet orbiting Proxima
Centauri;][]{2016Natur.536..437A} lack determinations of their bulk
densities $\varrho_{\rm p}$. Thus, the HZ terrestrial planets known
to date are not amenable to making clear statements on their
habitability.  The K2 \citep{2014PASP..126..398H}, TESS
\citep{2015JATIS...1a4003R}, and PLATO \citep{2018haex.bookE..86R}
missions are bound to provide tens of Earths and Super-Earths in the
HZ around bright M dwarfs and solar-type stars for which the estimates
of their bulk densities $\varrho_{\rm p}$ might be obtained in
principle, but atmospheric characterization for the latter sample
might be beyond the capabilities of JWST and the ground-based
telescopes with very large aperture diameters. The nearest stars to
the Sun are thus the most natural reservoir for the identification of
potentially habitable rocky planets that might be characterized via a
combination of high-dispersion spectroscopy and high-contrast imaging
with the ground-based telescopes \citep{Snellen+15} or via
coronagraphic or interferometric observations in space
\citep{Leger15}.

Unlike the Doppler and transit methods, astrometry alone can determine
reliably and precisely the true mass and three-dimensional orbital
geometry of an exoplanet, which are fundamental inputs to models of
planetary evolution, biosignature identification, and habitability.
By determining the times, angular separation and position angle at
periastron and apoastron passage, exquisitely precise astrometric
position measurements will allow the prediction of where and when a
planet will be at its brightest (and even the likelihood of a transit
event), thus (a) crucially helping in the optimization of direct
imaging observations and (b) relaxing important model degeneracies in
predictions of the planetary phase function in terms of orbit
geometry, companion mass, system age, orbital phase, cloud cover,
scattering mechanisms, and degree of polarization
\citep[e.g.][]{Madhusudhan&Burrows12}.  \emph{Only a high precision
  astrometric mission's observations will have the potential to 1)
  discover most of the potentially habitable planets around the
  nearest stars to the Sun, 2) directly measure their masses and
  system architectures, and 3) provide the most complete target list
  and vastly improve the efficiency of detection of potential habitats
  for complex exo-life with the next generation of space telescopes
  and ground-based very large aperture telescopes.}

\subsubsection{Fundamental Program} 

\begin{figure*}[tb]
\centering
\includegraphics[width=0.65 \hsize, trim = 0cm 0cm 0cm 1cm, clip]
{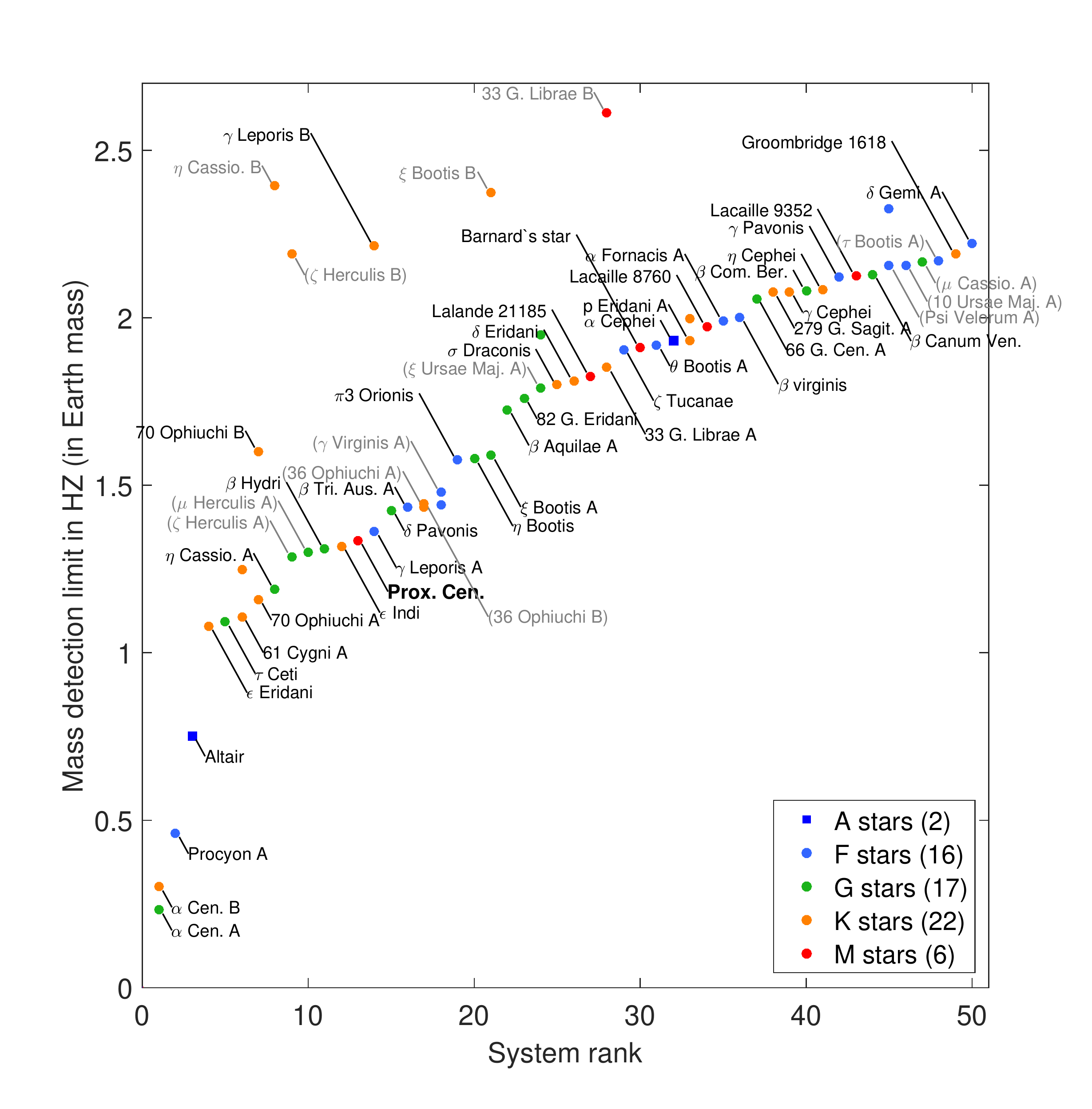}
 \caption{Minimum masses of planets that can be detected at the
  center of the HZ of their star for the 63 best nearby A, F, G, K, M
  target systems. The target systems (either single or binary stars),
  are ranked from left to right with increasing minimum detectable
  mass in HZ around the primary system component, assuming equal
  observing time per system. Thus for binary stars, A and B components
  are aligned vertically, as they belong to the same system therefore
  they share the same rank. When the A and B mass thresholds are close
  the name is usually not explicitly written down to avoid
  overcrowding. B components that have mass thresholds above 2.2
  $M_\oplus$ are named in gray and binaries that are estimated too
  close for follow-up spectroscopy are named in gray and in
  parenthesis. These binaries are expected to be only part of the
  secondary science program (planet formation around binaries). The
  star sample that is best for astrometry is similar to that of the
  best stars for spectroscopy in the visible, or in thermal IR (see
  text for explanations). Earths and Super-Earths with
  $M_{\rm p} \geq 1.5$ $M_\oplus$ can be detected and characterized
  (actual mass and full orbit) around 22 stars. All Super-Earths with
  $M_{\rm p} < 2.2$ $M_\oplus$ can be detected and characterized
  around 59 stars.  }
\label{fig:exoplan1}
\end{figure*}

\begin{figure*}[tb]
\centering
\includegraphics[width=0.95\hsize]{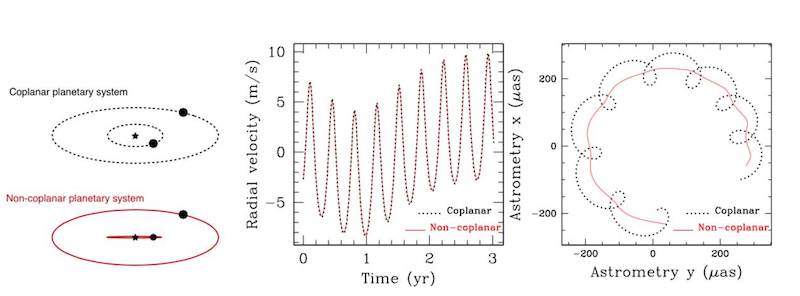}
 \caption{An example where astrometry breaks the degeneracy. Two
  simulated planetary systems are around a solar-type star at 10 pc,
  with two Jupiter-like planets at 0.5 and 2.5 AU (\textit{left}). One
  is co-planar (\textit{dotted black line}), the other has a mutual
  inclination of 30$^\circ$ (\textit{full red line}). The two
  corresponding RV curves are shown (\textit{middle}), as well as the
  two astrometric ones (\textit{right}). Curves are identical in the
  former case, but clearly separated in the latter revealing the
  inclined orbits.}
\label{fig:exoplan2}
\end{figure*}

Surgical single-point positional precision measurements in pointed,
differential astrometric mode ($<1\mu$as), could exploit a high
precision astrometric mission's unique capability to search for the
nearest Earth-like planets to the Sun. The amplitude $\alpha$ of the
astrometric motion of a star due to an orbiting planet is (in
microarcseconds):
\begin{equation}
  \alpha = 3\left(\frac{M_{\rm p}}{M_\oplus}\right)
  \left(\frac{a_{\rm p}}{1\,\mathrm{AU}}\right)
  \left(\frac{M_\star}{M_\odot}\right)^{-1}
  \left(\frac{D}{1\,\mathrm{pc}}\right)^{-1}\,\,\mu\mathrm{as}
\end{equation}
where $M_\star$ is the stellar mass, $M_{\rm p}$ is the mass of the
planet, $a_{\rm p}$ is the semi-major axis of the orbit of the planet,
and $D$ is the distance to the star. For a terrestrial planet in the
HZ of a nearby sun-like star, a typical value is 0.3 $\mu$as (an Earth
at 1.0 AU of a Sun, at 10 pc).  This very small motion (the size of a
coin thickness on the Moon as measured from the Earth) will be
accessible to a high precision astrometric instrument by measuring the
differential motion of the star with respect to far-away reference
sources.

A core exoplanet program could be comprised of 63 of the nearest A, F,
G, K, and M stars (Fig. \ref{fig:exoplan1}). Many of them are found in
binary and multiple systems. Binary stars are compelling for a high
precision astrometry space mission for a number of reasons.  They are easier
targets than single stars. % For close Sun-like binaries, the magnitude
% of both components is brighter than $V=9$ mag, which is the equivalent
% magnitude of a typical reference star field composed of 6 $V=11$ mag
% stars.

Furthermore, as the photon noise from the reference stars is the
dominant factor of the error budget, the accuracy for binaries
increases faster with telescope staring time than around single
stars. For binaries, the reference stars only need to provide the plate
scale and the reference direction of the local frame, the origin point
coordinates are constrained by the secondary/primary component of the
binary.  Finally, when observing a binary, the astrometry on both
components is obtained simultaneously: the staring time is only spent
once as both components are within the same field of view (FoV). These
two effects combined cause the observation of stars in binary systems
to be much more efficient (as measured in $\mbox{$\mu$as} \times
\mbox{h}^{-1/2}$) than that of single stars.

We further stress that the complete census of small and nearby planets
around solar-type stars is unique to high-precision astrometry.  On
the one hand, Sun-like stars have typical activity levels producing
Doppler noise of $\sim1$ m/s (or larger), which is still 10 times the
signal expected from an Earth-analog \citep{2011arXiv1107.5325L}.
High precision space astrometry will be almost insensitive to the
disturbances (spots, plages) due to stellar activity, having typical
activity-induced astrometric signals with amplitude below 0.1 $\mu$as
\citep{2011A&A...528L...9L}.

For the full sample of the nearest stars considered in Fig.~
\ref{fig:exoplan1} we achieve sensitivity (at the $6\sigma$ level) to
planets with $M_p \leq 3$ $M_\oplus$. If we consider
$\eta_\uplus\sim10\%$, for the sample of 63 stars closest to our Solar
System we thus expect to detect $\sim6$ HZ terrestrial planets. Of
these, 5 will be amenable for further spectroscopic characterization
of their atmospheres\footnote{One target is a binary which is too
  close for follow-up spectroscopy}.  A high precision astrometry
mission could perform the measurements of the relevant stars and make
a thorough census (95\% completeness) of these planets by using less
than 10\% of a four-year mission.  As indicated above, this program
will also be valuable for understanding planetary diversity, the
architecture of planetary systems (2-d information plus Kepler's laws,
results in 3-d knowledge) including the mutual inclination of the
orbits, a piece of information that is often missing in our
exploration of planetary systems.

\subsubsection{Additional Exoplanet Investigations}

A secondary program can help elucidate other important questions in
exoplanetary science.
\begin{enumerate}
\item {\bf Planetary systems in S-Type binary systems}. A high
  precision astrometry mission's performance for exoplanet detection
  around nearby binaries will be of crucial importance in revealing
  planet formation in stellar systems, the environment in which
  roughly half of main-sequence stars are born. The discovery of giant
  planets in binaries has sparked a string of theoretical studies,
  aimed at understanding how planets can form and evolve in highly
  perturbed environments \citep{2015pes..book..309T}. Giant planets
  around one component of a binary (S-type orbits) have often been
  found in orbits very close to theoretical stability limits
  \citep[e.g.][]{2004AIPC..713..269H, 2011CeMDA.111...29T,
    2016AN....337..300S}, and as for most of the binary targets the HZ
  of each component is stable, finding other and smaller bodies in
  their HZs is a real possibility. The contribution of a high
  precision astrometric mission could be decisive for these ongoing
  studies, by allowing the exploration of a crucial range of
  exoplanetary architectures in binaries.
\item {\bf Follow-up of known Doppler systems}. Another unique use of
  a high precision astrometry mission will be the study of
  non-transiting, low-mass mul\-ti\-ple-planet systems that have
  already been detected with Radial Velocities (RV). High precision
  astrometry will confirm or refute controversial detections, remove
  the $\sin i$ ambiguity and measure actual planetary
  masses. Furthermore, it will directly determine mutual inclination
  angles, which are critical to study (i) the habitability of
  exoplanets in multiple systems, since they modify the orientation of
  the spin axes and hence the way the climates change across time
  \citep[e.g.][]{1993Natur.361..608L,2013MNRAS.428.1673B,2014MNRAS.444.1873A}
  and (ii) the dynamical evolution history of multiple systems, as
  e.g. coplanar orbits are indicative of smooth evolution, while large
  mutual inclinations and eccentricities point toward episodes of
  strong interactions, such as planet-planet
  scattering. Fig.~\ref{fig:exoplan2} illustrates a case where
  degeneracy in RV can be removed by astrometry.  Using the proper
  motion difference technique or diagnostics representing 'excess'
  residuals to a single-star fit, there are a few Gaia-based results
  worth mentioning, such as mass constraints on the cool Super-Earth
  orbiting Proxima Centauri \citep{2020A&A...635L..14K}, the inferred
  true mass for HD\,114762b \citep{2019A&A...632L...9K}, and the first
  high-quality measurement of highly mutually inclined orbits in the
  Pi Mensae system \citep{2020A&A...642A..31D, 2020MNRAS.497.2096X}.
\item {\bf Planetary systems on and off the main sequen\-ce}.  Gaia
  will be able to detect thousands of giant planetary companions
  around stars of all ages (including pre- and post-main-sequence),
  spectral type, chemical abundance, and multiplicity with results
  expected in the DR4 and DR5 data releases
  \citep{2018haex.bookE..81S, 2018A&A...614A..30R,
    2020AJ....160...16S}.
  % Gaia has the potential to detect thousands of giant planetary
  % companions around stars of all ages (including pre- and
  % post-main-sequence), spectral type, chemical abundance, and
  % multiplicity
  % \citep{2008A&A...482..699C, 2014MNRAS.437..497S,
  % 2014ApJ...797...14P,2015MNRAS.447..287S}.
  A high precision astrometry mission could cherry-pick from Gaia
  discoveries and identify systems amenable to follow-up to search for
  additional low-mass components in such systems, particularly in the
  regime of stellar parameters difficult for radial velocity work like
  early spectral types, young ages, very low metallicity, white
  dwarfs. Some of the systems selected might also contain transiting
  companions identified by TESS and PLATO (and possibly even Gaia
  itself), or planets directly imaged by SPHERE on the VLT or European
  Extremely Large Telescope.
\item {\bf Terrestrial planets around Brown Dwarfs}. So far, among the
  few planetary mass objects that have been associated with brown dwarf
  (BD) hosts using direct imaging and microlensing techniques, only
  one is likely to be a low-mass planet \citep[][and references
  therein)]{2015ApJ...812...47U}.  However, there are both
  observational
  \citep{2008ApJ...681L..29S,2012ApJ...761L..20R,2014ApJ...791...20R}
  as well as theoretical
  \citep{2007MNRAS.381.1597P,2013ApJ...774L...4M} reasons to believe
  that such systems could also be frequent around BDs. The recent
  identification of a trio of short-period Earth-size planets
  transiting a nearby star with a mass only $\sim10\%$ more massive
  than the hydrogen-burn\-ing limit \citep{2016Natur.533..221G} is a
  tantalizing element in this direction. In its all-sky survey, Gaia
  will observe thousands of ultra-cool dwarfs in the backyard of the
  Sun with sufficient astrometric precision to reveal any orbiting
  companions with masses as low as that of Jupiter
  \citep{2014MmSAI..85..643S}. A high precision astrometry mission
  could push detection limits of companions down to terrestrial
  mass. If the occurrence rate of $P\leq1.3$ d, Earth-sized planets
  around BDs is $\eta=27\%$ as suggested by
  \citet{2017MNRAS.464.2687H} based on extrapolations from transit
  detections around late M dwarfs, the high precision measurements,
  probing for the first time a much larger range of separations with
  respect to transit surveys with sensitivity to low-mass planets,
  will unveil a potentially large number of such companions, and place
  the very first upper limits on their occurrence rates in case of
  null detection.
\item \textbf{Astrometric effect of disks.}  As pointed out by
  \citet{2016A&A...592A..39K}, the photocenter of a star+disc system
  will have an elliptic motion for asymmetric discs.  The latter is
  likely to mimic a planet or to perburb the characteriscs of an
  existing planet. Additional measurements (e.g.\ infrared flux) will
  be necessary to disentangle the disk asymmetry from a real planet.

\end{enumerate}
%%%%%%%%%%%%%%%%%%%%%%%%%%%%%%%%%%%%%%%%%%%%%%%%%%%%%%%%%%%%%%%%%%%%%
\subsection{Compact objects}
\label{sec:compact-objects}

% On Sept. 25, John Tomsick did significant editing of this entire
% section to reduce the length.  One place where more reduction may be
% possible is the caption of Figure 5.15, which was only slightly
% reduced.  Another outstanding item is to replace Figure 5.16 with a
% new version (see comments below)
%

\subsubsection{X-ray Binaries}
The brightest Galactic X-ray sources are accreting compact objects in
binary systems.  Precise optical astrometry of these X-ray binaries
provides a unique opportunity to obtain quantities which are very
difficult to obtain otherwise.  In particular, it is possible to
determine the distances to the systems via parallax measurements and
the masses of the compact objects by detecting orbital motion to
measure the binary inclination and the mass function.  With a high
precision astrometric mission, distance measurements are feasible for
$>$50 X-ray binaries\footnote{with 2\,000 hours of observation}, and
orbital measurements will be obtained for dozens of systems.  This
will revolutionize the studies of X-ray binaries in several ways: here
we discuss goals for neutron stars (NSs), including constraining their
equation of state (EoS), and for black holes (BHs).

\begin{figure}[t]
\centering
\includegraphics[width=0.8\hsize]{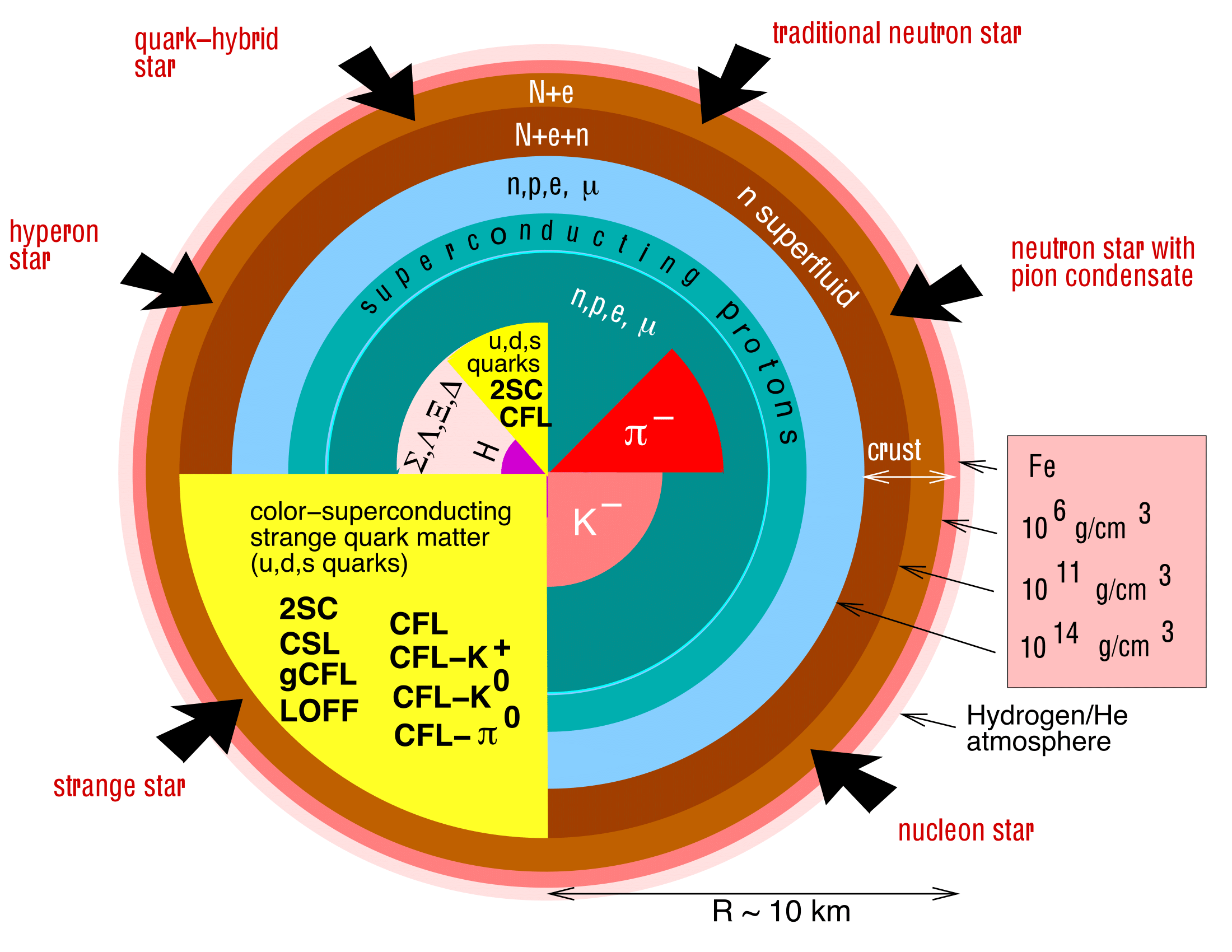}
 \caption{Sketch of the different existing possibilities for the
  internal structure of a neutron star.  Figure courtesy of
  \citet{Weber_2001}.}
\label{fig:nsstructure}
\end{figure}

Matter in the NS interior is compressed to densities exceeding
those in the center of atomic nuclei, opening the possibility to probe
the nature of the strong interaction under conditions dramatically
different from those in terrestrial experiments and to determine the
NS composition.  NSs might be composed of nucleons only;
% replacement
strange baryons (hyperons) or mesons might appear in the core or even
deconfined quark matter, forming then a hybrid star with a quark
matter core and hadronic matter outer layers; or of pure strange quark
matter (a quark star). 
% only, of strange
% baryons (hyperons) or mesons in the core with nucleons outside (a
% hybrid star), or of pure strange quark matter (a quark star). 
A sketch of the different possibilities is given in
Fig.~\ref{fig:nsstructure}. Via the equation of state (EoS), matter
properties determine the star's radius for a given mass. In
particular, since general relativity limits the mass for a given EoS,
the observation of a massive NS can exclude EoS models. Presently, the
main constraint stems from the measurements of very massive NSs in
radio pulsar/white dwarf systems which have been reported with high
precision \citep{Demorest2010, Antoniadis2013, 2018ApJS..235...37A, 2020NatAs...4...72C}.

\begin{figure*}[htb]
    \centering
    \hfill
    \includegraphics[width=0.35\hsize]{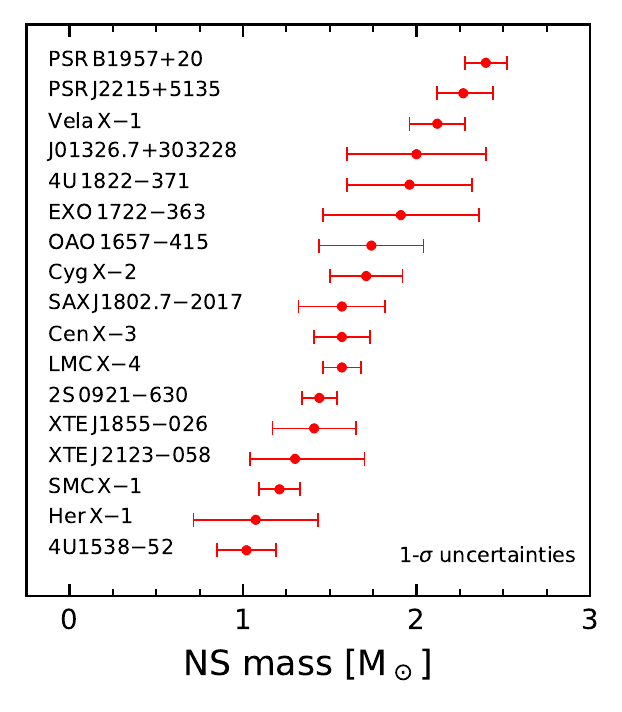}
    \hfill
    \includegraphics[width=0.45\hsize]{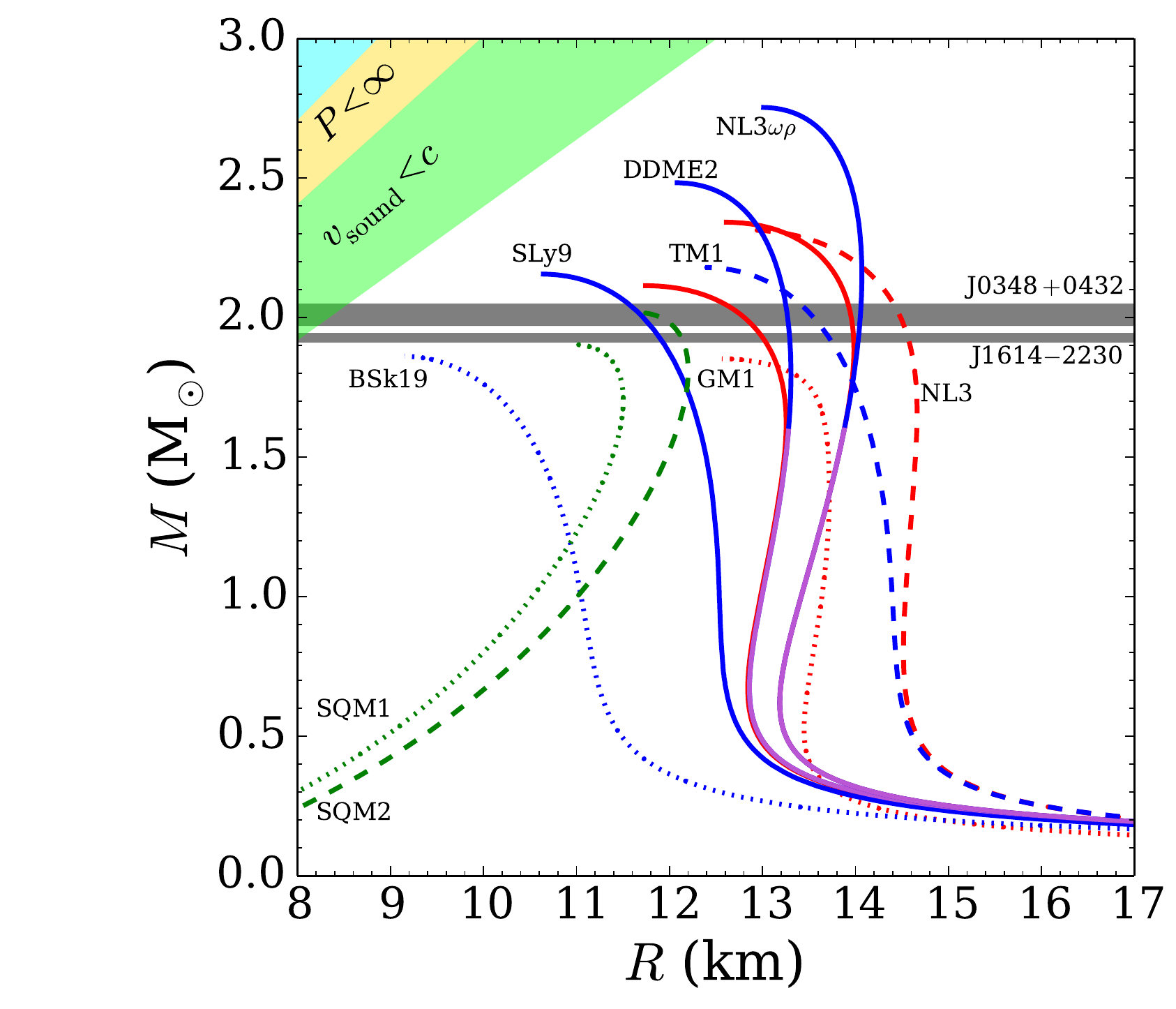}
    \hfill~\\
    \caption{Left: Neutron star mass measurements in X-ray
      binaries, update from \citet{Lattimer2005},
      \texttt{http://stellarcollapse.org}.  Right: $M$-$R$ relation
      for different EoS models (adapted from \citet{Fortin2016}): NS
      with a purely nucleonic core (in blue), with a core containing
      hyperons at high density (in red), and pure strange quark stars
      (in green). The horizontal grey bars indicate the masses of PSR
      J1614$-$2230 and PSR J0348+0432. The models indicated by dotted
      or dashed lines are either not compatible with NS masses or
      nuclear physics constraints. Note that a transition to matter
      containing hyperons is not excluded by present constraints.}
\label{fig:nsmasses}
\end{figure*}

The key to constraining the NS EoS is to measure the masses and radii
of NSs.  While masses have been measured for a number of X-ray binary
and radio pulsar binary systems \citep[e.g., ][]{Lattimer2016,
  Ozel2016}, the errors on the mass measurements for most X-ray
binaries are large (see Fig.~\ref{fig:nsmasses}, left).  The ultimate
constraint on the EoS will be a determination of radius and mass of
the same object, and a small number of such objects might be
sufficient to pin down the entire EoS (e.g. \citet{Ozel2009}), see
Fig.~\ref{fig:nsmasses} (right), where several $M$-$R$ relations for
different EoSs are shown.  Current techniques to determine radii rely
on spectroscopic measurements of accreting neutron stars, either in
quiescence \citep{Heinke2014} or during thermonuclear (type I) X-ray
bursts \citep{Ozel2016}, and also timing observations of surface
inhomogeneities of rotating NSs \citep{Miller2016, Haensel2016}.

A high precision astrometric mission will contribute by obtaining
precise mass constraints with orbital measurements \citep{Tomsick2010}
and by improving distance measurements.  Distances must be known
accurately to determine the NS radii.  For that purpose, new high
precision data can be combined with existing and future X-ray data,
e.g., from Athena, which is scheduled as the second large-class (L2)
mission in ESA’s Cosmic Vision.  The Athena Science Working Group on
the endpoints of stellar evolution has observations of quiescent
neutron star X-ray binaries to determine the NS EoS as its first
science goal; however, their target list is restricted to systems that
are in globular clusters.  A high precision astrometric mission will
enable distance measurements for many more NS X-ray binaries, allowing
Athena to expand their target list.

Other techniques for constraining the NS EoS might also be possible in
the future:
%replacement
detecting redshifted absorption lines; determining the NS moment of
inertia of systems like the double pulsar J0737$-$3039; and more
detections of gravitational wave (GW) emission by LIGO, Virgo or Kagra
from the inspiral of a binary neutron star merger like for GW170817
\citep{Abbott2017}.
GWs from the post-merger phase could strongly constrain the EoS, too.
% detecting redshifted absorption lines; determining the
% moment of inertia of the double pulsar J0737$-$3039;
% % replacement
% and more detections of gravitational wave (GW) emission from the
% insprial of a binary neutron star merger are expected after GW170817
% \citep{Abbott2017}.
% GWs from the post-merger phase could strongly constrain the EoS, too. 
% % and the detection
% % of gravitational wave emission from the inspiral of a NS-NS merger
% % \citep{Abbott2017}.
However, the mass and distance measurements that
a high precision astrometric mission will obtain use techniques that
are already well-established, providing the most certain opportunity
for greatly increasing the numbers of NSs with mass or radius
determinations.

%%%%%%%%%%%%%%%%%%%%%%%%%%%%%%%%%%%%%%%%%%%%%%%%
%%%%%%%%%%%%%%%%%%%%%%%%%%%%%%%%%%%%%%%%%%%%%%%%%%%

%%%%%%%%%%%%%%%%%%%%%%%%%%%%%%%%%%%%%%%%%%%%%%%%%%%%%
%%%%%%%%%%%%%%%%%%%%%%%%%%%%%%%%%%%%%%%%%%%%%%%%%%%%%

In addition to the goal of constraining the NS EoS, NS masses are also
relevant to NS formation and binary evolution.  Current evolutionary
scenarios predict that the amount of matter accreted, even during
long-lived X-ray binary phases, is small compared to the NS mass.
This means that the NS mass distribution is mainly determined by birth
masses.  Determining the masses of NSs in X-ray binaries, therefore,
also provides a test of current accretion models and evolutionary
scenarios, including the creation of the NSs in supernovae.

BHs are, according to the theory of general relativity, remarkably
simple objects.  They are fully described by just two parameters,
their mass and their spin. Precise masses are available for very few
BHs in X-ray binaries. The recent detection of gravitational waves
\citep{Abbott2016c} found in the binary BH mergers
\citep{Abbott2016b,2019PhRvX...9c1040A} show that they have, on
average, higher masses and probably lower spins than the BHs in X-ray
binaries.  These measurements are difficult to explain based on our
understanding of stellar evolution and the fate of massive
stars. Although BHs leave few clues about their origin, one more
parameter that can be determined is the proper motion of BHs in X-ray
binaries. Measurements of proper motions provides information about
their birthplaces and formation. It includes whether they were
produced in a supernova (or hypernova) or whether it is possible for
massive stars to collapse directly to BHs.  A few BH X-ray binaries
have proper motion measurements (e.g., \citet{Mirabel2001}), but this
number will rise dramatically with the astrometry measurements that a
high precision astrometry mission will provide.

Currently, the cutting edge of research in BH X-ray binaries involves
constraining BH spins, including the rate of spin and the orientation
of the spin axis.  Techniques for determining the rate of spin include
measuring the relativistic broadening of the fluorescent iron
$K_\alpha$ line in the X-ray emission and the study of the thermal
continuum X-ray spectra \citep{Remillard2006, Miller2007}. Concerning
the direction of their spin axes, there is evidence that the standard
assumption of alignment between the BH spin and orbital angular
momentum axes is incorrect in some, if not many, cases
\citep{Maccarone2002, Tomsick2014, Walton2016}, likely requiring a
warped accretion disc. Theoretical studies show that such
misalignments should be common \citep{King2016}. However, binary
inclination measurements rely on modeling the ellipsoidal modulations
seen in the optical light curves \citep{Orosz2011}, which is subject
to systematic uncertainties, and a high precision astrometry mission
will be able to provide direct measurements of orbital inclination for
many of the BH X-ray binaries that show evidence for misalignments and
warped discs.

\subsubsection{Astrometric microlensing}
\label{sec:Compact objects in the GC}

\begin{figure*}[htb]
  \centering
  \includegraphics[width=.98 \hsize]{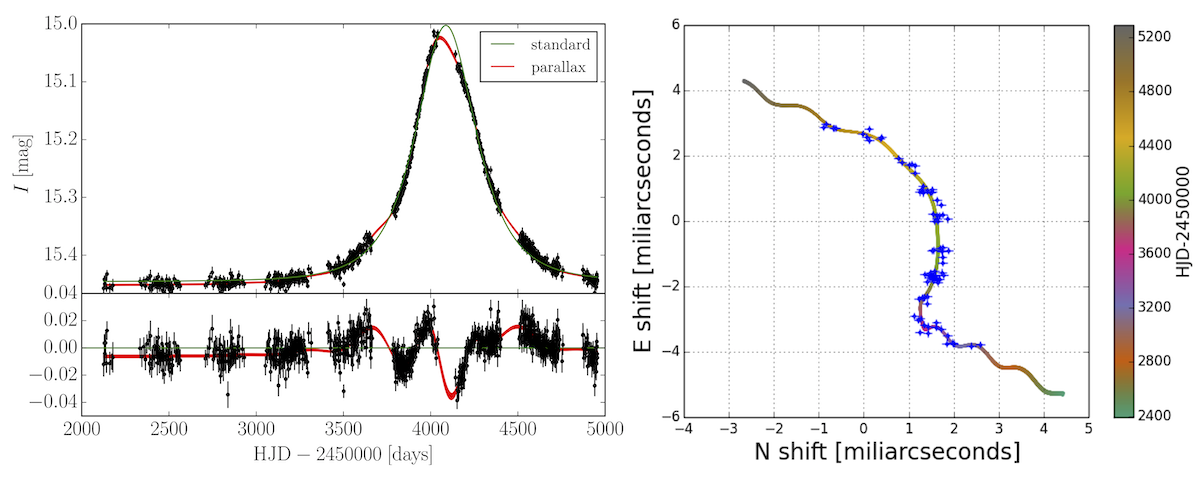}
  \caption{Microlensing event, OGLE3-ULENS-PAR-02, the best
    candidate for a $\sim$10M$_\odot$ single black hole. Left:
    photometric data from OGLE-III survey from 2001-2008. Parallax
    model alone can only provide mass measurement accuracy of
    50-100$\%$. Right: simulated astrometric microlensing path for a
    similar event if observed with Theia, a high-precision astrometry
    mission. Combining this mission's superb astrometric accuracy with
    long-term photometric data will yield mass measurements of black
    holes and other dark compact object to 1$\%$ even at faint
    magnitudes.}
    \label{fig:microlensing}
\end{figure*}

In 1986 Bohdan Paczy{\'n}ski \citep{Paczynski1986} proposed a new
method for finding compact dark objects, via photometric gravitational
microlensing. This technique relies on continuous monitoring of
millions of stars in order to spot the temporal brightening due to
space-time curvature caused by the presence and motion of a dark
massive object. Microlensing reveals itself also in astrometry, since
the centre of light of both unresolved images (separated by $\sim$1
mas) changes its position while the relative brightness of the images
changes in the course of the event. Astrometric time-series at sub-mas
precision over the course of a couple of years will provide
measurement of the size of the Einstein Ring, which combined with the
photometric light curve, will directly yield the lens distance and
mass. Most microlensing events are detected by large-scale surveys,
e.g., OGLE and, in future possibly also the Rubin Observatory
(previously known as the LSST).  At typical
brightness of V=19-20mag only a high-precision astrometry mission will
be capable of providing good-enough astrometric follow-up of
photometrically detected microlensing events
(Fig.~\ref{fig:microlensing}). Among 2000 events found every year, at
least a couple should have a black hole as the lens, for which the
mass measurement via astrometric microlensing will be possible.
\let\figmicrolensig=\ref{fig:microlensing}

% Map of Theia's sweetest world 
\begin{figure*}[t] 
\centering
\includegraphics[width=0.75\hsize]{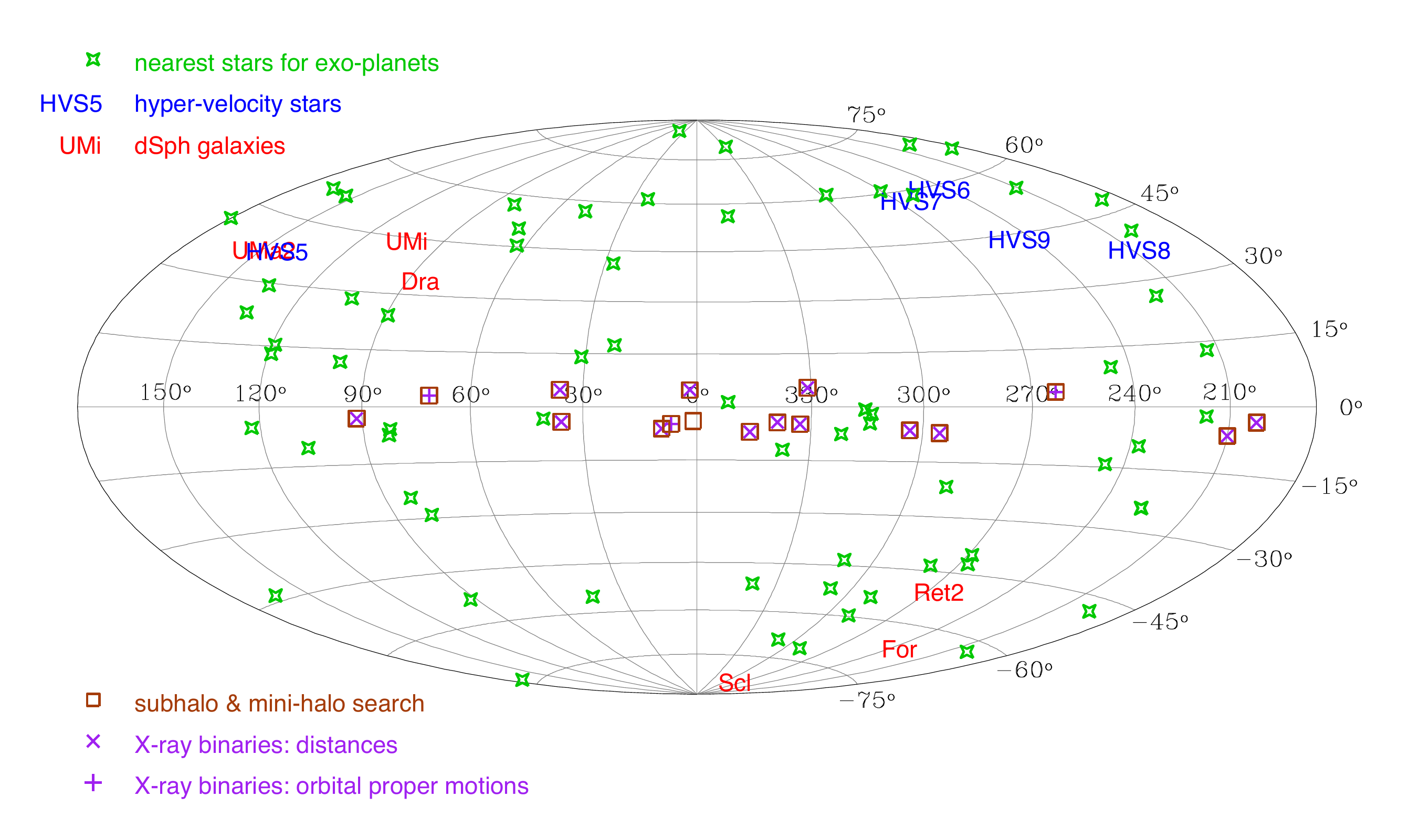}
\caption{Sky map of the targets considered for observations with a
  high precision astrometric mission.}
\label{fig:skyplot}
\end{figure*}

Detection of isolated black holes and a complete census of masses of
stellar remnants will for the first time allow for a robust
verification of theoretical predictions of stellar
evolution. Additionally, it will yield a mass distribution of lensing
stars as well as hosts of planets detected via microlensing.

%%%%%%%%%%%%%%%%%%%%%%%%%%%%%%%%%%%%%%%%%%%%%%%%%%%%%%%%%%%%%%%%%%%%%
\subsection{Cosmic distance ladder}
\label{sec:cosm-dist-ladd}

The measurement of cosmological distances has revolutionized modern
cosmology and will continue to be a major pathway to explore the
physics of the early Universe. The age of the Universe ($H_0^{-1}$) is
a key measurement in non-standard DM scenarios.  Its exact value is
currently strong\-ly debated, with a number of scientific papers
pointing at discrepancies between measurement methods at the
2-3$\sigma$ level. But the most serious tension with a discrepancy at
the 3-4 $\sigma$ level appears between CMB estimates
($H_0 = 67.8 \pm 0.9$\,km/s/Mpc) or for that matter BAO results from
the SDSS-III DR12 data combined with SNIa which indicate
$H_0 = 67.3 \pm 1.0$\,km/s/ Mpc \citep[see][]{Alam2016b} and
measurements based on Cephe\-ids and SNIa \citep{2016ApJ...826...56R}
giving $H_0 = 73.24 \pm 1.74$\,km/s/Mpc.

The tension between the methods can be due to unknown sources of
systematics, to degeneracies between cosmological parameters, or to
new physics \citep[e.g.][]{Karwal&Kamionkowski16}. It is therefore of
crucial importance to consider methods capable of measuring $H_0$ with
no or little sensitivity to other cosmological
parameters. Uncertainties can be drastically reduced by measuring time
delays (TD) in gravitationally lensed quasars \citep{Refsdal64}, as
this technique only relies on well-known physics that is General
Relativity. With enough statistics, and good modeling of the mass
distribution in the lensing galaxy, TD measurements can lead to
percent-level accuracy on $H_0$, independently of any other
cosmological probe \citep[e.g. ][]{Bonvin2016a, Suyu2013, Suyu2014}.
In practice, TDs can be measured by following the photometric
variations in the images of lensed quasars. As the optical paths to
the quasar images have different lengths and they intersect the lens
plane at different impact parameters, the wavefronts along each of
these paths reach the observer at different times. Hence the notion of
TD.

Significant improvements in lens modeling, combined with long-term
lens monitoring, should allow measuring $H_0$ at the percent
level. The H0LiCOW program ($H_0$ Lenses in COSMOGRAIL's Wellspring),
which focuses on improving the detailed modeling of the lens galaxy
and of the mass along the line of the sight to the background quasar,
led to $H_0 = 73.3 ^{ +1.7}_{-1.8}$\,km/s/Mpc (that is 2.4\% precision)
% $H_0 = 71.9 \pm 2.7$\,km/s/Mpc (that is 3.8\% precision)
in a flat $\Lambda$CMD Universe by using deep HST imaging, Keck
spectroscopy and AO imaging and wide field Subaru imaging \citep{
  Bonvin2016a, Rusu2016, Sluse2016, Suyu2016, Wong2016,
  2020MNRAS.498.1420W}. This value is in excellent agreement with the
most recent measurements using the distance ladder (though in tension
with the CMB measurements from Planck) but still lacks precision.

By performing photometric measurements with the required sensitivity
and no interruption, the combination of a high precision astrometric
mission and excellent modeling of the lens galaxy, will enable
measurement of $H_0$ at the percent level and remove any possible
degeneracies between $H_0$ and other cosmological parameters. This
will open up new avenues to test the nature of DM. An alternative
technique consists of using trigonometric parallaxes. This is the only
(non-statistical and model-independent) direct measurement method and
the foundation of the distance scale.  A high precision astrometric
mission has the potential to extend the ``standard candles'' - the more
distant pulsating variables: Cepheids, RR Lyrae, Miras and also
Stellar Twin stars - well beyond the reach of Gaia.

These distance measurements can be transferred to nearby galaxies
allowing us to convert observable quantities, such as angular size and
flux, into physical qualities such as energy and
luminosity. Importantly, these distances scale linearly with $H_0$,
which gives the temporal and spatial scale of the Universe. With this
improved knowledge, we will then be able to better understand the
structure and evolution of both our own and more distant galaxies, and
the {\it age} of our Universe.

\subsection{Position of the science targets in the sky}
\label{sec:posit-science-tar}

The different targets considered for observations with a high
precision astrometry mission have been located in
Fig.~\ref{fig:skyplot} on a sky map.

%%%%%%%%%%%%%%%%%%%%%%%%%%%%%%%%%%%%%%%%%%%%%%%%%%%%%%%%%%%%%%%%%%%%%
%%%%%%%%%%%%%%%%%%%%%%%%%%%%%%%%%%%%%%%%%%%%%%%%%%%%%%%%%%%%%%%%%%%%%
\section{Possible space mission}
\label{sec:poss-space-miss}

Several mission profiles have been considered in the last few years
focused on differential astrometry, for instance NEAT, micro-NEAT and
Theia. Additional new differential astrometry mission configurations
adapted with technological innovations will certainly be envisioned to
pursue accurate measurements of the extremely small motions required
by the science cases in this White Paper.

%%%%%%%%%%%%%%%%%%%%%%%%%%%%%%%%%%%%%%%%%%%%%%%%%%%%%%%%%%%%%%%%%%%%%
\subsection{Scientific requirements}
\label{sec:scient-requ}

%%%%%%%%%%%%%%%%%%  CELINE
To address the science described in this white paper, a high precision
astrometry mission should stare towards:
\begin{itemize}
\setlength\itemsep{0em} 
\item dwarf spheroidal galaxies to probe their dark matter inner
  structure;
\item hyper-velocity stars to probe the triaxiality of the halo, the
  existence of compact minihalo objects and the time delay of quasars;
\item the Galactic Disc, to probe DM subhalos and compact minihalo
  objects; 
\item star systems in the vicinity of the Sun, to find the nearest
  potentially habitable terrestrial planets; 
\item known X-ray binaries hosting neutron stars or black holes.
\end{itemize}

\begin{figure}[t]
\centering
\includegraphics[width=0.8\hsize,trim = 0.5cm 0cm 0cm 0cm, clip]{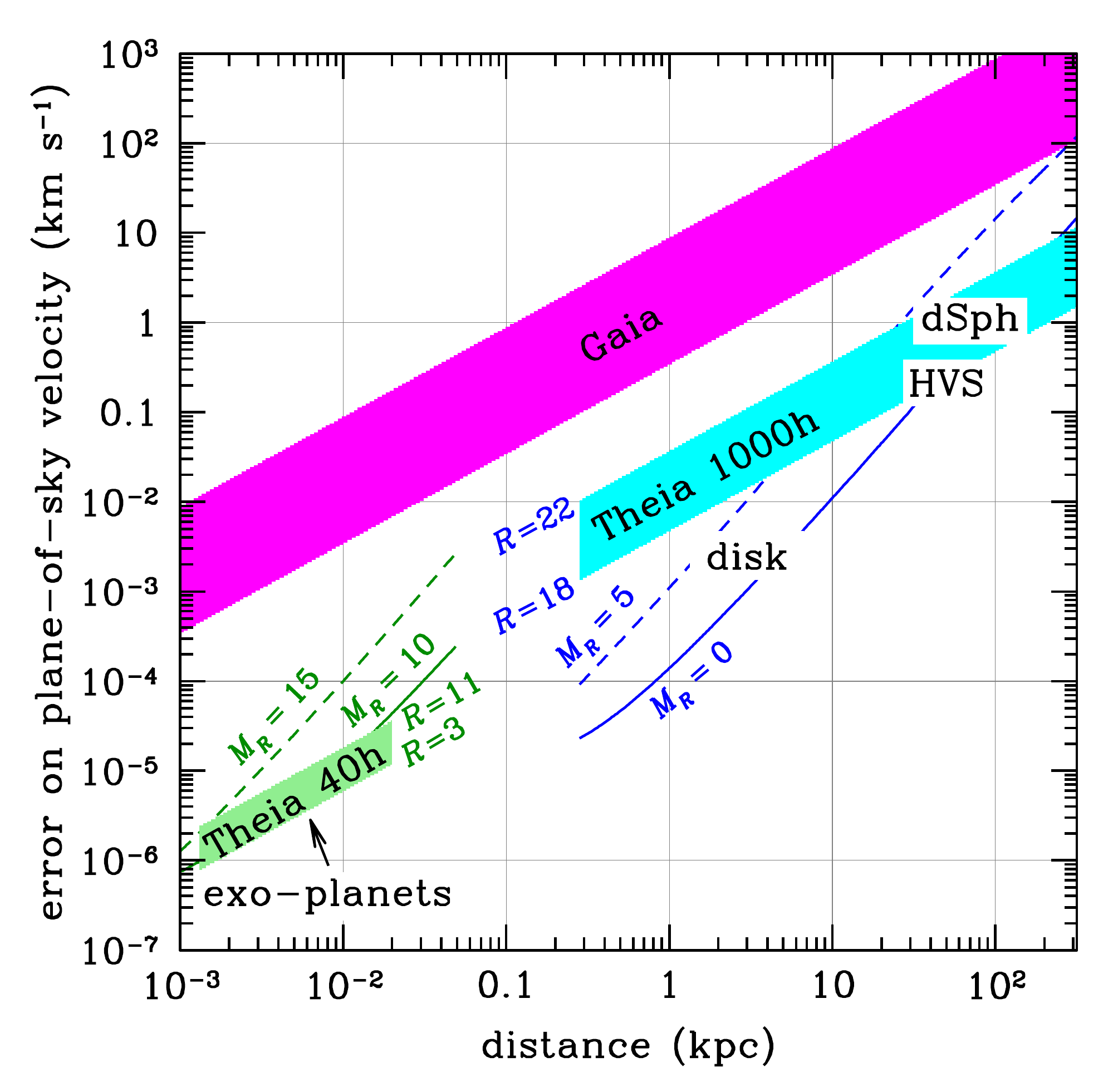}
 \caption{Expected plane-of-sky velocity errors from a high
  precision astrometry mission's proper motions as a function of
  distance from Earth. These errors respectively correspond to 40 and
  1000 cumulative hours of exposures for exo\-planets (\textit{green})
  and more distant objects (\textit{cyan} and \textit{blue}), during a
  4 year interval for observations, including the systematic limit
  from calibration on Gaia reference stars. The expected precision for
  specific objects are highlighted. The accuracy for the 5-year Gaia
  mission is shown in \textit{magenta}.}
\label{fig:vposerrvsDist}
\end{figure}

\begin{figure*}[t] 
  \centering
  % \hfill
  % \includegraphics[width=0.42\hsize,clip]{parallaxMissionComparison_v4_forProposal.pdf}
  % \hfill
  % \includegraphics[width=0.42\hsize,clip]{PropMotMissionComparison_v5_forProposal.pdf}
  % \hfill~\\
  \hfill
  \includegraphics[width=0.85\hsize,clip]{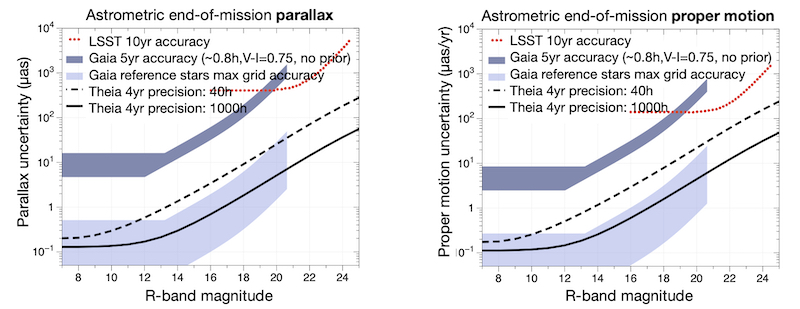}
  \hfill~\\  \caption{Estimated RMS precision on a high precision astrometry
    mission relative parallax (\textit{left}, for ecliptic latitude
    $0^\circ$) and proper motion (\textit{right}) in the
    $R$-band. Also shown for comparison are the estimated accuracies
    for 10~years LSST \citep{2009arXiv0912.0201L} as well as the
    5-year nominal Gaia mission \citep{deBruijne+15} (vertical spread
    caused by position on the sky, star color, and bright-star
    observing conditions). Small-scale spatial correlations
    ($<$1$^\circ$) between Gaia reference sources will limit the
    maximum reachable absolute parallax and proper motion calibration
    for a high precision astrometry mission
    \citep{2012A&A...543A..14H, 2012A&A...543A..15H}, indicated by the
    light blue band for a range of assumed spatial correlations
    \citep[expected to be much below $r=0.5$\%;][]{2009IAU...261.1703H}
    as a function of reference star magnitude. Bright stars ($V<13$) and
    low star-density regions will have the highest correlations.}
\label{fig:astroComparisons}
\end{figure*}

\begin{table*}[t]
  \small
  \caption{Summary of science cases with most stringent
    performance requirements set in each case. Figures are based on a
    4 year mission, thermal stabilisation ($+$slew time) is assumed to
    take 30$\%$ of the mission time. 
    \label{tab:tech.summary.science}}
  \centering
  \begin{tabular}{lrrrrr} 
  %\begin{tabular}{lrrrrrr} 
    \hline 
Program 
& Used
& Mission 
& Number of
& Benchmark target 
& EoM precision \\
%& Ref.\ frame \\ 
& time (h)
& fraction 
&  objects per field 
& $R$ mag (and range) 
& (at ref.\ mag.) \\
%& accuracy $\mu^{N\to\infty}_0$\\ 
    \hline 
    \hline %\\
    Dark Matter & 
    17\,000 &
    0.69 &
    10$^2$--10$^5$ &
    20 (14--22) &
    10 $\mu$as  \\ 
    %& $2-5\,\mu$as/yr\\
    $\&$ compact objects  
    \\
    \hline
    Nearby Earth-like planets           
    & 3\,500 &
    0.14 &
    $<$20 &
    5 (1--18)  &
    0.15 $\mu$as  \\ 
    $\&$ follow-up \\
    %&(none)    \\
    \hline
    Open observatory & 
    4\,000 &
    0.17 &
    10-10$^5$ &

    6 (1-22) &
    1.0 $\mu$as  \\
    %&$2-5\,\mu$as/yr
    %\\
    \hline 

    Overall requirements& 
    24 \,500 &
    1.00 &
    10$^1$-10$^5$ &
    6 (1-22) &
    0.15-10 $\mu$as \\  
    %&$2-5\,\mu$as/yr
    %\\
    \hline 
  \end{tabular}
\end{table*}
For a targeted mission, the objects of interest must be sampled
throughout the lifetime of the mission. After re-pointing the
telescope and while waiting for stabilization, photometric surveys,
e.g. for measurements of $H_0$ using lensed quasar time delays could
be performed, thus optimizing the mission scientific
throughput. Fig.~\ref{fig:skyplot} shows a sample sky map with
potential targets.

As illustrated in Fig.~\ref{fig:vposerrvsDist}, high precision
astrometric missions could measure the plane-of-sky velocities of the
faintest objects in the local Universe, with errors as small as a few
mm/s in the case of the hosts of Earth-mass exoplanets in the
habitable zone of nearby stars, a few m/s for stars in the Milky Way
disc, i.e. for kinematical searches for DM sub-halos, micro-lensing
searches for ultra-compact minihaloes, and for the companions of
neutron stars and black holes in X-ray binaries, 200m/s for
hyper-velocity stars whose line of sight velocities are typically
$>500$\,km/s, and finally 1\,km/s for $R=20$\, mag stars for dwarf
spheroidal galaxies.

A mission concept with an expected Theia-like astrometric precision,
as shown in Fig.~\ref{fig:astroComparisons}, surpasses what will be
achieved by other approved space astrometric surveys and ground
surveys, thus unlocking science cases that are still unreachable.

Table \ref{tab:tech.summary.science} summarizes the science cases with
the most stringent performance requirements. To cover the science
questions from this White Paper, any mission concept must be flexible,
allowing for observing modes covering a wide flux dynamical
range. This requires the concepts to cope with \emph{Deep Field
  Modes}, aimed towards objects such as dwarf galaxies, and
\emph{Bright Star Modes}, focused on the study of planetary systems
around nearby stars.

%%%%%%%%%%%%%%%%%%%%%%%%%%%%%%%%%%%%%%%%%%%%%%%%%%%%%%%%%%%%%%%%%%%%%
\subsection{Example of a medium-size mission}
\label{sec:prop-scient-instr}

\begin{figure*}[t]
  \centering
  \includegraphics[width=0.9\hsize]{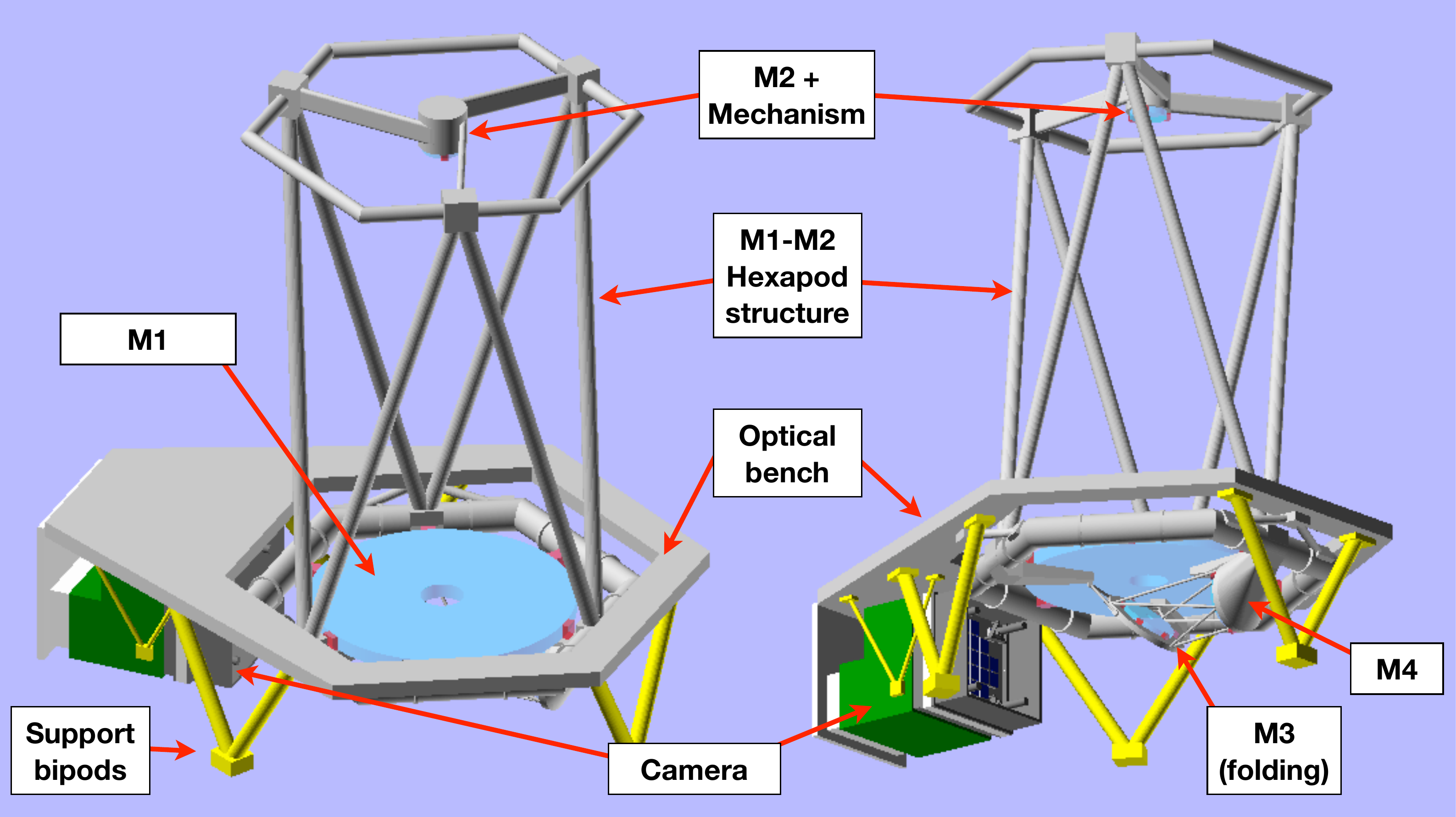}  
  \caption{Overall layout of the Theia Payload Module concept. Volume
    is estimated at $1.6\times1.9\times2.2$m$^3$.}
  \label{fig:theiam5-plmconcept}
\end{figure*}

The Payload Module (PLM) of a high precision astrometric mission must
be simple. It is essentially composed of four subsystems: telescope,
camera, focal plane array metrology and telescope metrology. In the
case of the Theia/M5 concept, they were designed applying heritage
from space missions and concepts like Gaia, HST/FGS, SIM, NEAT
(proposed for the ESA M3 opportunity),
Theia (proposed for the ESA M4 opportunity), and Euclid.

However, achieving microarcsecond differential astrometric precision
requires the control of all effects that can impact the determination
of the relative positions of the point spread function. The typical
apparent size of an unresolved star corresponds to 0.2~arcseconds for
a 0.8~m telescope operating in visible wavelengths. The challenge is
therefore to control systematic effects to the level of 1 part per
200\,000. The precision of relative position determination in the
Focal Plane Array (FPA) depends on i) the photon noise, which can be
either dominated by the target or by the reference stars; ii) the
geometrical stability of the instrument, iii) the stability of the
optical aberrations, and iv) the variation of the detector quantum
efficiency between pixels. The control of these effects impairs other
missions that otherwise could perform microarcsecond differential
astrometry measurements, like HST, Kepler, the Roman Space Observatory
(previously known as WFIRST), or Euclid, posing
fundamental limits to their astrometric accuracy. All these effects
must be taken into account in any high precision differential
astrometry mission concept.

To address the challenges and fulfil the requirements from section
\ref{sec:scient-requ}, two different possible concepts can be
investigated. A NEAT-like mission consisting of a formation flight
configuration \citep{2012ExA....34..385M} or an Euclid-like
mission,\footnote{Euclid red book:
  \url{http://sci.esa.int/euclid/48983-euclid}
  \url{-definition-study-report-esa-sre-2011-12}.} but with a single
focal plane and additional metrology subsystems. Both concepts are
based on adopting a long focal length, diffraction-limited telescope,
and additional metrological control of the focal plane array. The
proposed Theia/M5 mission concept was the result of a trade-off
analysis between both concepts.

\subsubsection{Telescope concept}

\begin{figure}[t]
\centering
\includegraphics[width = 0.9\hsize]{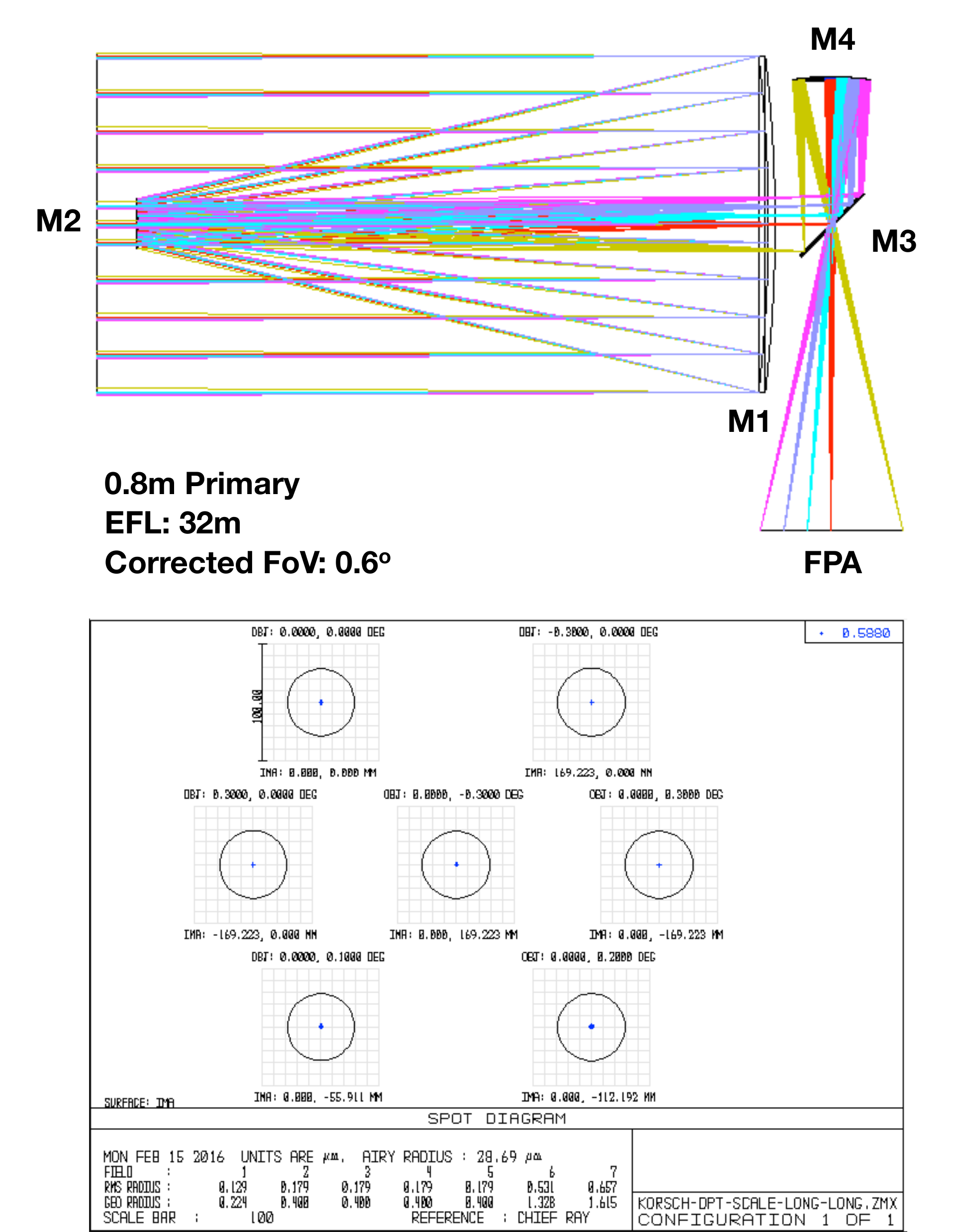}\\*[1em]
\caption{On-axis Korsch TMA option. Ray-tracing and spot diagrams for
  the entire FoV. This design was adopted as the baseline for the
  Theia/M5 proposal. EFL is Effective Focal Length.}
\vspace{-0.4cm}
\label{fig:theiam5-onaxis-korsch}
\end{figure}

The Theia PLM concept consists of a single Three Mirror Anastigmatic
(TMA) telescope with a single focal plane (see
Fig.~\ref{fig:theiam5-plmconcept}) covering a $0.5^\circ$
field-of-view with a mosaic of detectors. To monitor the mosaic
geometry and its quantum efficiency, the PLM includes a focal plane
metrology subsystem, while to monitor the telescope geometry, a
dedicated telescope metrology subsystem is used.

To reach sub-microarcsecond differential astrometry a
diffraction-limited telescope, with all aberrations controlled, is
necessary. A trade-off analysis was performed between different
optical designs, which resulted in two optical concepts that could
fulfil all requirements. Both are based on a Korsch TMA telescope;
one is an on-axis solution while the second is an off-axis
telescope. In both cases only three of the mirrors are powered
mirrors. While the on-axis solution adopts a single folding mirror,
the off-axis solution adopts two folding mirrors. The on-axis design
was the Theia/M5 baseline (Fig.~\ref{fig:theiam5-onaxis-korsch}). More
recently, however, studies from NASA/JPL show that a customized and
corrected Rit\-chey-Chrétien can reach 5 $\mu$as over a $0.5^\circ$ FoV,
which even if not capable of addressing habitable exoplanet science
cases, would provide a valuable instrument for Dark Matter studies.

\begin{figure}[t]
\centering %
\includegraphics[width = \hsize]{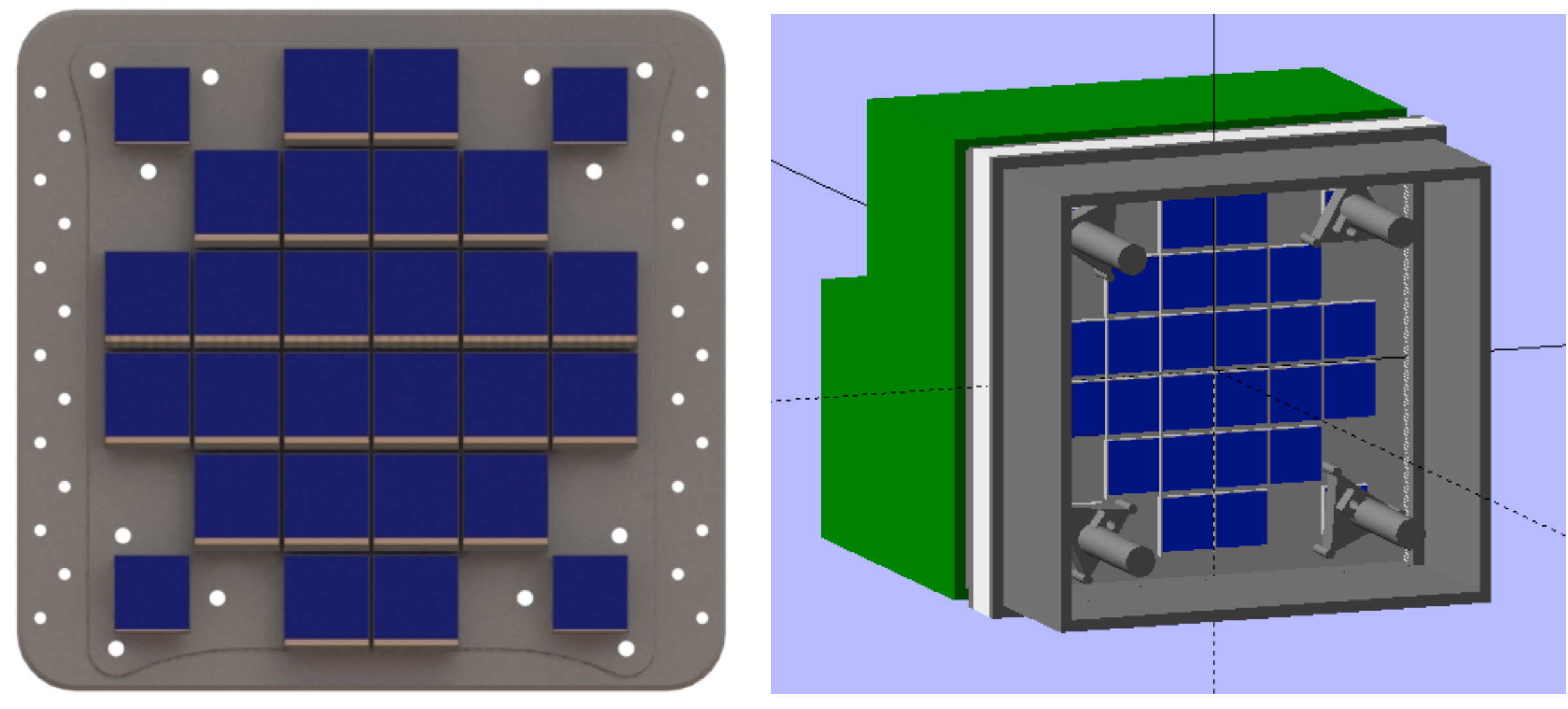}\\*[1em]
\vspace{-0.4cm}
\caption{Concept for the Theia/M5 Camera. Left: concept for the FPA detector
  plate. Right: overall view of the camera concept.}
\label{fig:theiam5-fpaconcept}
\end{figure}

To achieve the precision by centroiding as many stars as possible, a
mosaic of detectors (in principle CCD or CMOS) must be assembled on
the focal plane (Fig.~\ref{fig:theiam5-fpaconcept}). The detectors
must feature small pixels ($\sim 10\,\mu$m) and well controlled
systematic errors along the lifetime of the mission. Detailed in-orbit
calibration of the focal plane and detector geometry and response must
be monitored, and in the Theia concept this is addressed via a
dedicated laser metrology (see Sect.~\ref{sec:metrology}).

\subsubsection{Mission configuration and profile}
\label{sec:miss-conf-prof}

\begin{figure*}[p]
  \centering
  \includegraphics[width=0.98\hsize]{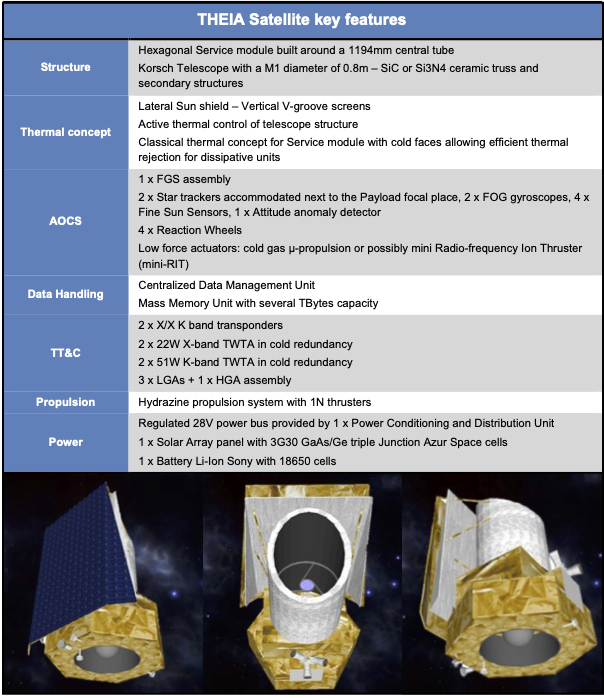}
  \caption{Proposed Theia satellite concept (Thales Alenia Space).
  FGS: Fine Guidance Sensor; FOG: Fiber Optics Gyroscope;  AOCS:
  Attitude and Orbit Control System,TT\&C: Telemetry, Tracking \&
  Control;  TWTA: Travelling Wave Tube amplifier Assembly; LGA: Low Gain Antenna;
  HGA: High Gain Antenna.}
  \label{fig:mission.sc.concept}
\end{figure*}

\begin{table}[t]
  \centering
  \caption{Theia's mission main characteristics.}
  \smallskip
  \includegraphics[width=\hsize]{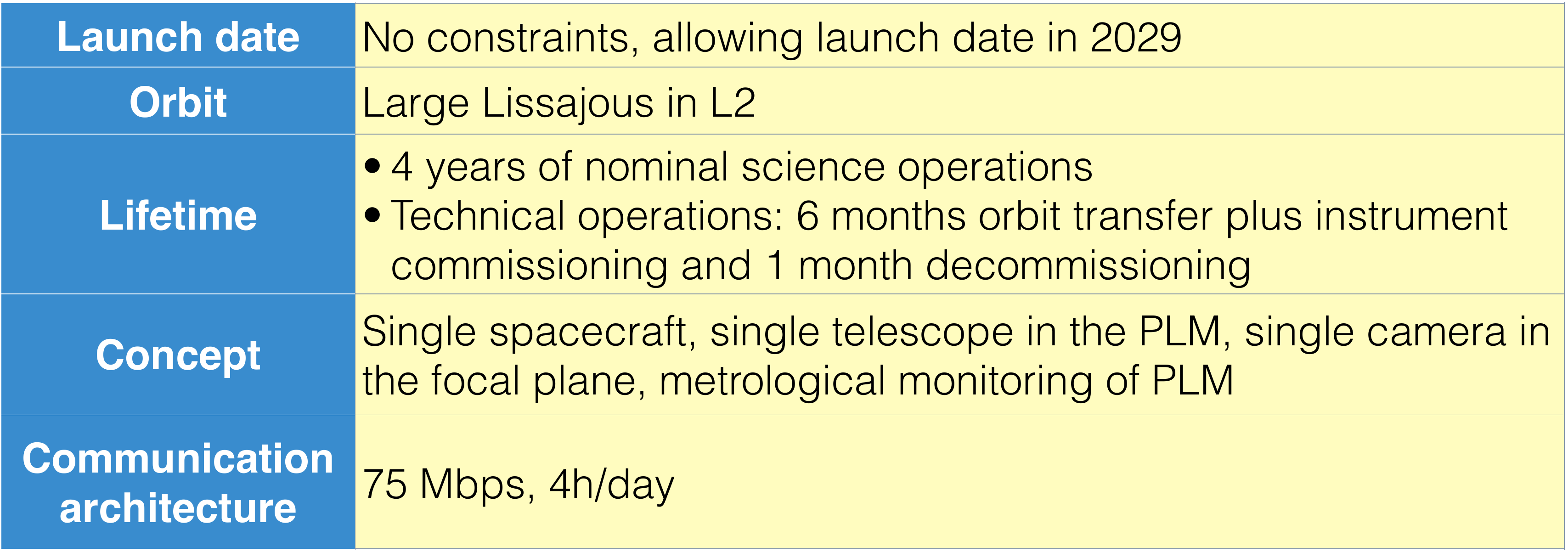}
  \vspace{-0.4cm}
  \label{tab:mission-main-characteristics}
\end{table}

The time baseline (Table~\ref{tab:mission-main-characteristics}) to
properly investigate the science topics of this White Paper would be
at least 4 years, including time devoted to orbit maintenance. A total
of approximately 6 months has been estimated for the orbit transfer
including the spacecraft and instrument commissioning. This estimate
is made from the total of $\sim 35\,000$ h corresponding to the total
time for the scientific program (Table \ref{tab:tech.summary.science})
and considering that about 15 min per slew will be dedicated to
reconfiguration and station-keeping, while thermal stabilization time
is in addition to the slew time.

Some instrument key features of the Theia concept are presented in
Fig.~\ref{fig:mission.sc.concept}. The concept is inspired by the
Euclid service module with a downscaled size to minimize mass and
improve mechanical properties. Similar to the Euclid and Herschel
satellites, Theia's Korsch telescope is accommodated on top of the
service module in a vertical position leading to a spacecraft height
of about 5m. This concept optimizes the payload size.

%%%%%%%%%%%%%%%%%%%%%%%%%%%%%%%%%%%%%%%%%%%%%%%%%%%%%%%%%%%%%%%%%%%%%
%%%%%%%%%%%%%%%%%%%%%%%%%%%%%%%%%%%%%%%%%%%%%%%%%%%%%%%%%%%%%%%%%%%%%
\section{Worldwide context of ground-based and space science}

Observations carried out with a mission dedicated to high precision
astrometry will add significant value and will benefit from a number
of other ground-based and space missions operating in the 2030s and
beyond, including ESA's Athena, PLATO, Euclid and Gaia missions, ESO's MICADO
and Gravity instruments, CTA, SKA, the NASA/ESA/CSA JWST and the Rubin
Observatory (previously known as LSST). For example:
\begin{itemize}
\item \textbf{JWST:} Estimates suggest that JWST will be able to
  detect Lyman Break galaxies with absolute magnitudes as faint as
  $M_{\rm{UV}} \sim -15$ at $z\sim7$ \citep{2013MNRAS.434.1486D},
  corresponding to halo masses of about $10^{9.5} \ M_{\odot}$. The
  combination of a high precision astrometry mission and JWST's
  observations will enable unambiguous tests of DM.
\item \textbf{PLATO:} this mission will look at planetary transits and
  star oscillations in two fields (each covering 2250 deg$^2$), for
  2-3 years each, in host stars brighter than 16 mag.  PLATO high
  cadence continuous monitoring of its target stars will provide
  information on the internal structure of the stars, allowing
  determination of their stellar ages and masses. A high precision
  astrometry mission will benefit from PLATO characterization of many
  of the astrometry mission's core star samples.  For close PLATO
  stars where transits were observed this astrometry mission can
  measure additional inclined planets.
\item \textbf{SKA:} SKA aims to use radio signals to look for building
  blocks of life (e.g. amino acids) in Earth-sized planets
  \citep{2015aska.confE.120Z, 2015aska.confE.115H}.  A high precision
  astrometry will identify target planets from their astrometric
  ``wobble'' that can be followed-up spectroscopically with the
  SKA. Furthermore, SKA aims to use its immensely fast sky coverage to
  detect transients \citep{2015aska.confE..51F}, such as supernovae and
  gamma-ray bursts. With its precise astrometry, Theia will help study
  the specific locations of such events in stellar clusters.
\item \textbf{CTA: } The Cherenkov Telescope Array (CTA) in the
  Northern and Southern Hemispheres will carry out measurements of the
  gamma-ray flux with almost complete sky coverage and unprecedented
  energy and angular resolution, in the $\sim$ 0.02-100\,TeV energy
  range \citep{2011ExA....32..193A}. The sub-micro\-arc-sec\-ond
  performance of a high precision astrometry mission allow us
  investigating the so-called J-factor which corresponds to the
  brightness of the gamma-ray flux in dwarf galaxies and thus
  determines the prime candidates for CTA's observations. CTA also
  aims at observing star forming systems over six orders of magnitude
  in formation rate, to measure the fraction of interacting cosmic
  rays as a function of the star-formation rate. By combining high
  precision astrometry and CTA measurements, we will better understand
  the relative importance of cosmic rays and DM in places where
  star-formation is important. Furthermore, a small number of black
  hole and neutron star binary systems in our Galaxy are known to emit
  gamma rays. The mechanism by which the particle acceleration is
  achieved is not well understood. The sub-micro\-arcsecond
  performance of a high precision astrometry mission will al\-low us
  to probe the velocity structure of the nearby gam\-ma-ray bright
  radio galaxies of NGC\,1275, IC\,310, M\,87, and Cen\,A, which
  combined with CTA's observations will enable important astrophysics
  breakthroughs.
\end{itemize}

%%%%%%%%%%%%%%%%%%%%%%%%%%%%%%%%%%%%%%%%%%%%%%%%%%%%%%%%%%%%%%%%%%%%%
%%%%%%%%%%%%%%%%%%%%%%%%%%%%%%%%%%%%%%%%%%%%%%%%%%%%%%%%%%%%%%%%%%%%%
\section{Technology challenges for high precision astrometry}
\label{sec:cont-other-miss}

\subsection{Spacecraft technology and cost}

There have been several propositions for a space mission dedicated to
high precision astrometry: a 6 meter baseline visible interferometer
on a single satellite like SIM or SIM-Lite
\citep{2008SPIE.7013E..4TG}; a single mirror off-axis parabola
1\,m-diameter telescope based on two spacecraft, one carrying the
telescope mirror and the other the focal plane like the NEAT telescope
\citep{2012ExA....34..385M}; or a single-mirror telescope like Theia
\citep{2016SPIE.9904E..2FM, Boehm2017}. The variety of the concepts
shows that there are areas of progress on spacecraft technologies,
especially concerning formation flying, actively-controlled
  large structure interferometers.

One interesting potential solution to be considered is the nanosat
technology and the cost reduction that is linked to it. There is a
huge cost difference between cubesats ($< 10$\,M\euro{}) and an ESA
M-class mission ($400-500$\,M\euro{}) or NASA MIDEX/Discovery mission
($300-500$\,M\$). The cubesat technology has matured and many hundreds
are launched every year. That technology has now crept into micro-sats
that are up to 200\,kg and spacecraft bus of this category are now
$<5$\,M\euro{}, while only a few years ago they were
$\sim40$\,M\euro{}.  Because of their low cost and the high number of
flying satellites, this technology has now demonstrated 5 year typical
lifetime, comparable to a more expensive traditional spacecraft. In
any case, all the price scales will change between now and the epoch
when Voyage 2050 will be implemented and that includes flying heavier
payloads because of the decrease of launch costs \citep{Jones2018}.
  
\subsection{Detection}

Presently, two detector technologies are used: CCD or CMOS. CMOS
detectors present a high quantum efficiency over a large visible
spectral band that can also reach infrared wavelengths depending on
the sensitive layer. CMOS detectors also have programmable readout
modes, faster read\-out, lower pow\-er, better radiation hardness, and
the ability to put specialized processing within each pixel. On the
other hand there are many known detector systematics, even for
ad\-van\-ced detectors like the Teledyne H4RG10. The main challenging
effects are the following: fluence-dependent PSF, correlated read
noise, inhomogeneity in electric field lines and persistence effects
\citep[e.g.][]{Simms2009}. All mission proposals so far were based on
CCD technology, but detector evolution will certainly take place on
the context of any mission concept to answer the challenges being
posed by the Voyage 2050 White Papers.

If a Theia-like mission is selected for the 2040’s, detector
technology might be different from anything we have in place
nowadays. The main requirements are small pixels, low read-out noise (RON)
on large format focal plane and mastering intrapixels effects in order
to reach the highest precision astrometry. It should be noticed that
the development of European detector technology for low-RON and
large-format IR and visible detector matrices, like the Alfa detector
that ESA is undertaking with Lynred, is of high interest for our
science cases.
  
\subsection{Metrology}
\label{sec:metrology}

Traditionally systematic errors have been the major challenge for
$\mu$as-level astrometry from space. Astrometric accuracy has a lot in
common with photometric accuracy, and the technology development that
proceeded the Kepler mission demonstrated $\sim 10^{-5}$ relative
photometry.  Similar advances have been made in detector calibration
for astrometry \citep{2016SPIE.9904E..5GC}.  Photons from stars carry
the astrometric information at exquisite precision, systematic errors
are imparted when those photons strike the telescope optics and also
when they are detected by the focal plane array. The calibration of
optical field distortion using reference stars is a technique that is
perhaps a century old and used on ground and space-based telescopes.

Metrology laser-feed optical fibers placed at the back of the nearest
mirror to the detectors can be used to monitor distortions of the
focal plane array, and to allow the associated systematic errors to be
corrected \citep{2016SPIE.9904E..5GC}. Such detector calibration at
$10^{-6}$ pixel levels should be continued. In addition to measuring
the FPA physical shape, the rest of the telescope needs monitoring to
control time-variable aberrations at sub $\mu$as level. Even at very
stable space environments such as L2, the telescope geometry is
expected to vary for different reasons: structural lattice
reorganization (such as the micro-clanks observed in ESA's Gaia mission), outgassing
and most importantly, thermo-elastic effects due to the necessary
variation of the Solar Aspect Angle during the mission for pointings
to the different science targets.

In the case of Theia, the telescope metrology subsystem to monitor
perturbations to the telescope geometry is based on a concept of a
series of simple and independent linear displacement interferometers
installed between the telescope mirrors and organized in a virtual
hexapod configuration. Existing space-based interferometers from TNO,
as the Gaia Basic Angle Monitor (BAM) are already capable of reaching
more precise measurements than those required by Theia/M5 -- BAM can
perform $\sim1.5$ pm optical path difference measurements
\citep{doi:10.1117/12.2026928}. A Thales telemeter developed for CNES
can reach $\sim100$ pm, and the Thales interferometer produced for the
MTG (Meteosat Third Generation) satellite can reach 1 nm per
measurement \citep{Scheidel2011} -- higher precisions can be reached
by averaging over many measurements.

For telescopes that do not have high stability levels, there are
some alternatives. One is the diffractive pupil concept that puts a
precision array of dots on the primary, which produces a regular
pattern of dots in the focal plane. One way to use the diffractive
pupil is to look at a very bright star (0 mag) and record the
diffraction pattern interspersed with observations of a much dimmer
target star ($\sim7$\,mag).  The diffractive pupil can also be used
during science observations, but when the target star is $\sim7$\,mag
photon noise of the diffracted light can be significantly higher than
the photon noise of the reference stars ($\sim11-14$\,mag).

\section*{Conclusion}

To solve fundamental questions like
\begin{itemize}
\item ``\emph{What is the nature of dark matter?}''
\item ``\emph{Are there habitable exo-Earths nearby?}''
\item ``\emph{What is the equation of state of matter in extreme environments}?''
\item ``\emph{Can we put direct constraints on cosmological models and dark energy parameters?}''
\end{itemize}
many branches of astronomy need to monitor the motion of faint objects
with significantly higher precision than what is accessible
today. Through ultra-precise micro\-arcsecond relative
  astrometry, a high precision astrometry space mission will address
  the large number of important open questions that have been detailed in
  this White Paper.

The scientific requirements points toward a space mission that is
relatively simple: a single telescope, with metrology subsystems and a
camera. Such a mission can fit as an M-class mission, or even at a
smaller mission class depending on the final accuracy which is desired.

Some technological challenges must be tackled and advanced: the
spacecraft, the focal plane detector and the metrology. We believe
that these challenges can be mastered well before 2050 and that they
will open the compelling scientific window of the faint objects in
motion.

\begin{acknowledgements}
%If you'd like to thank anyone, place your comments here and remove the percent signs.
  The authors would like to thank the researchers and engineers who
  are not co-authors of this paper but who have taken part and have
  brought their contribution to the proposed missions to ESA
  successive calls: NEAT (M3), micro-NEAT (S1), and Theia (M4 and
  M5). An extensive list of supporters for the science objectives is
  given in \cite{Boehm2017}. We thank also Arianna Gallo for her
  contribution in our investigation of the shape of the Milky Way dark
  matter halo and Krzysztof A.\ Rybicki who generated the
  plots from Fig.~\ref{fig:microlensing}.

  We are grateful to the anonymous referee who helped to improve the
  quality of the paper with his/her remarks.

  Concerning the funding of our work, we would like to acknowledge the
  support of many agencies or programs. %%
  R.B.\ acknowledges support from NASA's Virtual Planetary Laboratory lead
  team under cooperative agreements NNA\-13\-AA\-93A. % R. Barnes
  A.C.M.C.\ acknowledges support from CFisUC strategic project
  (UID/FIS/04564/2019).  %A.C.M Correia
  F.C.\ acknowledges support by the Swiss National Science Foundation (SNSF)
  and by the European Research Council (ERC) under the European Union’s
  Horizon 2020 research and innovation program (COSMICLENS: grant agreement
  No.~787886). % F. Courbin
  M.F.\ received support from Polish National Science Centre (NCN) under
  Grant No.~2017/26/D/ST9/00591. %M. Fortin
  M.F.\ gratefully acknowledge the support of the Swedish National Space
  Agency (DNR 65/19, 174/18). %M. Fridlund
  D.H.\ thanks the Swedish National Space Agency (SNSA/Rymdstyrelsen) for
  their support. % D. Hobbs
  A.M.\ thanks the Portugese Fundação para a Ciência e a Tecnologia (FCT)
  through the Strategic Programme UID/FIS/00099/2019 for
  CENTRA. % A. Moitinho de Almeida
  P.S.\ acknowledges support from the Australian Research Council under
  grant FT190100814. % P. Scott
  L.W.\ acknowledges support from the Polish NCN grants: Harmonia
  No.\ 2018/06\/M/ST9/00311 and Daina No.\ 2017/27/L/ST9/03221. % L. Wyrzykowski
  The OATo team acknowledges partial funding by the Italian Space Agency
  (ASI) under contracts 2014-025-R.1.2015 and 2018-24-HH.0, and by a grant
  from the Italian Ministry of Foreign Affairs and International
  Cooperation (ASTRA). % M. Gai
  A.C.\ and F.M.\ acknowledge support by the LabEx FOCUS ANR-11-LABX-0013.
  The work of C.J., X.L.\ and J.P.\ was supported by the Spanish Ministry of
  Science, Innovation and University (MICIU/FEDER, UE) through grants
  RTI2018-095076-B-C21, ESP2016-80079-C2-1-R, and the Institute of Cosmos
  Sciences University of Barcelona (ICCUB, Unidad de Excelencia ’Mar\'{\i}a
  de Maeztu’) through grants MDM-2014-0369 and
  CEX2019-000918-M. %C. Jordi, J. Portell, X. Luri
  A.K.-M., A.A., V.C., P.G., P.G., A.M.A., A.M., M.S.\ were supported by
  Funda\c{c}\~{a}o para a Ci\^{e}ncia e a Tecnologia, with grants reference
  UIDB/00099/ 2020 and SFRH/BSAB/142940/2018 (P.G.\ only). % P. Garcia et al.
  A.D. and L.O.\ also acknowledge partial support from the Italian
  Ministry of Education, University and Research (MIUR) under the
  Departments of Excellence grant L.232/2016, and from the INFN grant
  InDark. % A. Diafero et al.
  G.J.W. gratefully acknowledges support of an Emeritus Fellowship
  from The Leverhulme Trust. % G. White.
  E.V. is supported by Spanish grant PGC2018-101950-B-100.

  This research has made use of NASA’s Astrophysics Data System
  Bibliographic Services.  % Diafero add+++

\end{acknowledgements}


\bibliographystyle{aa}
\bibliography{EXPA_malbet_SI_Voyage2050}

\end{document}